\documentclass[]{aa}
\usepackage{graphicx}

\newcommand{\be}{ \begin {equation}}
\newcommand{\ee}{ \end {equation}}

\usepackage[utf8x]{inputenc}
\usepackage{amsmath}

\def\Xint#1{\mathchoice
   {\XXint\displaystyle\textstyle{#1}}%
   {\XXint\textstyle\scriptstyle{#1}}%
   {\XXint\scriptstyle\scriptscriptstyle{#1}}%
   {\XXint\scriptscriptstyle\scriptscriptstyle{#1}}%
   \!\int}
\def\XXint#1#2#3{{\setbox0=\hbox{$#1{#2#3}{\int}$}
     \vcenter{\hbox{$#2#3$}}\kern-.5\wd0}}

\def\dashint{\Xint-}

\begin{document}

\titlerunning{A retrograde  binary  embedded in an accretion disk}

\title  {The evolution of a binary in a retrograde circular orbit embedded in an accretion disk }

\author{
          P.B. Ivanov \inst{1,2}
        \and
        J.  C. B. Papaloizou \inst{2}
        \and
        S.-J. Paardekooper  \inst{3}
        \and 
        A.G. Polnarev \inst{3}
}
\institute{
Astro Space Centre, P. N. Lebedev Physical Institute, 84/32 Profsoyuznaya st., Moscow, 117997, Russia \\
\and
DAMTP, University of Cambridge, Wilberforce Road, Cambridge CB3 0WA, UK \\
          \and
        Astronomy Unit, Queen Mary University of London, Mile end Road, London,  E1 4NS, UK \\
 \\       \email{pbi20@cam.ac.uk}
}

\date{Draft Version \today} 

\abstract
{}
{Supermassive black hole binaries may form as a consequence of galaxy mergers. Both prograde and retrograde orbits have been proposed. We study  a binary of a small mass ratio, $q$, in a retrograde  orbit immersed in and interacting with  a gaseous accretion disk in order to estimate time scales for inward migration leading to coalescence and the accretion rate to the secondary component.}
{We employ  both semi-analytic methods and two dimensional numerical simulations, focusing on the case where the binary mass ratio is small but large enough to significantly perturb the disk.}
{We develop the theory of type I migration for  this case and   determine conditions for gap formation   finding  that  then  inward migration occurs on a  time scale equal to the time required for one half of the secondary mass to be accreted  through the unperturbed  disk, with accretion  onto the secondary playing only  a minor role. The semi-analytic and  fully numerical approaches are in good agreement, the  former being  applicable  over long time scales. }
{Inward migration induced by interaction with the disk alleviates the final parsec problem. Accretion onto the secondary does not significantly affect the orbital evolution, but may have observational consequences for high  accretion efficiency. The binary may  then appear as two sources of radiation rotating around each other. This study should be extended
to consider orbits with significant eccentricity and the effects of gravitational radiation  at small length scales.  Note too  that torques acting between a circumbinary  disk  and  a  retrograde binary  orbit may cause the mutual inclination to  increase  on a timescale that can be similar  to,  or smaller than  that for orbital evolution,  depending on detailed  parameters. This is also an aspect for future study. }

\keywords{Accretion disks: -binaries, Hydrodynamics, Galaxies: quasars: supermassive black holes, Planet-disk interactions}

\maketitle

\section{Introduction}

The  merger of two galaxies could lead to the formation of a supermassive binary black hole (SBBH) in the centre 
of  the newly formed more massive galaxy 
as a result of dynamical friction, (see e.g.   Komberg 1968, Begelman, Blanford \&  Rees 1980).
The coalescence of supermassive black holes results in a burst of gravitational radiation, which  can 
in principle  be detected  by planned space based gravitational antennae,  when originating from  cosmological  
distances, (see e.g. Grishchuk et al. 2001; Amaro-Seoane et al. 2013). 

The orbit of SBBH shrinks towards coalescence as a result of the operation of a number of mechanisms.
These include;  frictional interaction with a gaseous component that may reside in the form of an accretion disk,
gravitational interaction with the stars of a central star cluster, and the  emission of gravitational waves. 
It was noted by  Begelman, Blanford \&  Rees (1980)
that  orbital evolution of the SBBH induced by  gravitational radiation and gravitational interaction
with a star cluster  may not be efficient at intermediate scales of order of $0.01-1pc$ so that
it  may stall when its semi-major axis reaches these scales. This has been called the final parsec problem.

This problem   may be alleviated by taking into account the
interaction with a  gaseous  component of the system assumed to be in the form
of a  circumbinary accretion disk, see e.g. Ivanov, Papaloizou $\&$ Polnarev (1999), hereafter IPP, 
Gould $\&$ Rix (2000), and more recently e.g Cuadra et al. (2009), 
Haiman, Kocsis $\&$ Menou (2009), Lodato et at (2009), Rossi et 
al.  (2010), Farris et al.  (2011), Roedig et al.  (2012),
Kocsis, Haiman $\&$ Loeb (2012), Tanaka, Menou $\&$ Haiman (2012),
Hayasaki, Saito $\&$ Mineshige (2013), Rafikov (2013) and  Gold et al. (2013) and references therein. 
Apart from providing a possible solution to the final parsec problem,  the time variability of the accretion rate
supplied by an  accretion disk around a SBBH, peculiarities in emission spectra, etc. may serve
to indicate  its  presence (  e.g. Rieger $\&$ Mannheim 2000; Yu $\&$ Lu 2001; Armitage $\&$ Natarajan 2002,
Liu 2004; Lobanov $\&$ Roland 2005; Komossa 2006;
McFadyen $\&$ Milosavljevic 2008; Bogdanovic et al. 2008, 2009;  Montuori et al. 2011, 2012;  Sesana et al. 2012; 
Valtonen, Ciprini $\&$ Lehto 2012;  Burke-Spolaor 2013;
Ju et al. 2013; 
Farris et al. 2014;  McKernan et al. 2014;  Roedig et al., 2014  and references therein).
 In addition, the filling of the  central regions of the accretion disk after SBBH coalescence  under various circumstances,
 may also provide indirect observational evidence
 that the event  of SBBH  coalescence actually takes place ( see e.g.  Loeb 2007;  Shapiro 2010).

It has been  commonly assumed that the direction of motion of the disk gas and the orbital motion of the  SBBH  coincide.
However,  as noted by  Nixon, King $\&$ Pringle (2011) and Nixon et al. (2011),   this may not always be the case.  Since  the  direction 
of the binary orbital motion may  not be  correlated with the  initial direction of the gas motion within the disk.  
For example the system considered may  correspond to the inner regions of a disk galaxy which on large scales consists of  counter rotating gas and stars (see e.g. Corsini 2014 for a review) 
with  the gas containing a  relatively small part  of the total angular momentum content. 
In that case gas starting with extreme  orbital inclination relative to the stellar component  is  naturally expected to  settle into a counter rotating state (e.g. Thakar \& Ryden 1998).

Thus the SBBH  may be  aligned  with  orbital motion
being either  prograde or   retrograde  with respect to the direction of orbital motion of the disk gas.
 This can be understood as follows.
 Let us consider a SBBH consisting of  a primary with mass $M$ and a secondary with mass $M_p$, and assume, for simplicity, that 
the mass ratio $q$ is small, $q \ll 1$.  We  assume that there is an 
inclined thin circumbinary  accretion disk with a typical mass, $M_d$,  within the scale of the SBBH semi-major axis,  $a,$  which is  much 
smaller than $M_p.$
At distances much larger than $a$ the gravitational field of SBBH may be approximated by it's time average form, namely
 the  gravitational field  due to a  ring of mass $M_p$ and radius $a$.
 It was shown
by IPP that the disk must align into the binary plane on a  scale larger than $a$.
 Clearly,  when the mass of the secondary greatly exceeds the  disk mass  on this  alignment scale, and as here  only the time averaged potential of the binary matters, 
 the direction of alignment will  not depend on the direction of SBBH binary motion, and the disk may align with either 
prograde or retrograde   orbital motion, depending on which of these requires the inclination change of least magnitude (see King et al. 2005)
\footnote{Hydrodynamic interaction of the secondary with 
the accretion disk during the alignment process is discussed in e.g. Ivanov, Igumenshchev $\&$ Novikov
(1998).}.
In this paper we shall assume that the  configuration of the system and source of disk gas enables retrograde binary orbital motion with respect to the disk
to be set up.  Note, however, that after the initial  alignment process,  gravitational torques  exerted between
a  twisted  circumbinary disk   and  a slightly misaligned  binary orbit,    through which there is an accretion flow,  may tend to  overturn  the  orbital plane of the binary  on a
 long time scale when there is  counterrotation ( see e.g. Scheuer $\&$ Feiler 1996;  IPP; King et al.  2005;  Nixon, King $\&$ Pringle  2011). 
 The  timescale associated with this process is determined by the
magnitude  of the gravitational torque exerted by the  twisted accretion  disk on the binary.
 In the linear regime of small mutual inclinations,  this torque can be quite large 
when  the effective  Shakura-Sunyaev  $\alpha$ parameter  (Shakura \& Sunyaev 1973) governing the evolution of the twisted disk  is small. 
However,  once the  inclination  of the disk with respect
to the orbital plane at large distances is significant,  non-linear effects come into play and are expected to reduce the value of  the torque
(see e.g.  Ogilvie 1999). In this situation  we estimate  the timescale of the evolution of binary orbital plane 
 assuming  that the effective $\alpha \sim 1$ as was  done in IPP (see their equation (23)).  We discuss this estimate further in section \ref{twd}  showing  that the time scale  is expected
to be comparable to  or even slightly smaller than the timescale of evolution of the binary semi-major axis,  depending on the  disk  parameters and binary mass ratio. 
Thus,  depending on these,  the binary may be expected to evolve in the state of retrograde
rotation for  long  enough to undergo  significant changes to its orbit  and  in some cases 
 the effect of secular changes of its orientation may  become  important.   However, for simplicity,  in what follows we neglect
this possibility,  assuming that the planes of the binary and the disk are aligned. 

The evolution of a  SBBH immersed in the accretion disk and rotating in the retrograde sense has  recently been explored numerically  using the SPH method 
for a relatively large mass ratio $q \ge 0.1$  (see e.g.  Roedig $\&$ Sesana 2014).   In addition McKernan at al. (2014) have recently considered the opposite limit 
of an extremely small mass ratio $q=10^{-4}$ 
using the  grid based PENCIL CODE. 

In this Paper we present a simple semi-analytic  theory of a retrograde binary, with  mass ratio in the range we consider, immersed in and interacting with  a thin
 accretion disk together with the results of  two dimensional numerical  hydrodynamical simulations of representative systems.   
Unlike  the works mentioned above, we consider the case when the  mass ratio $q \le 2\cdot 10^{-2}$ is small but still large enough to significantly affect the
evolution of the disk surface density.  We consider both disks of formally infinite 
extent and disks with a finite outer boundary with various prescriptions for the  effective viscosity.
As two dimensional numerical simulations become problematic for disks with a large dynamic range that are likely
to be relevant for SBBH, we develop a  simpler applicable  one dimensional approach validating through
making comparisons with the two dimensional simulations.
For this first treatment we focus on  the simplest case of a  binary with a circular orbit interacting 
only with the disk  while briefly discussing features that should be tackled in future work such as orbital eccentricity and 
orbital evolution induced by gravitational radiation.

The plan of this Paper is as follows.
We give the basic  definitions and   set-up in section \ref{s1}
and   go on to describe a framework for  a  one dimensional model for calculating the  evolution of the disk  and binary orbit in section \ref{1Dmodel}.
In this model, the evolution of the disk surface density is governed  by a diffusion equation that incorporates
a  simple model of the angular momentum exchange 
 between the perturber and disk through gravitational scattering.
For small  mass ratios the perturber remains embedded in the disk,  producing a linear response  that  induces   type I migration.
The theory of this is presented  in section \ref{s2} and expressions for the migration rate given.
A comparison with migration rates  obtained from
two dimensional numerical simulations is then undertaken in section \ref{Esim}. The two approaches are found to be in good agreement.
Angular momentum exchange with the disk is expected  to result in gap formation for sufficiently massive 
perturbers.  This process  is studied in section \ref{Nonlin} using the one dimensional model set up in section \ref{s1}.
 Gap formation was  determined  to   occur for perturber mass ratios
exceeding an estimated value $\sim 1.57\delta^2,$ with $\delta$ being the disk aspect ratio. 
 Simple 1D modelling  of the gap surface density profile for the case of relatively large mass ratio is described in section \ref{s3}. 

As two dimensional  simulations are time consuming and impractical for disks with a very large dynamic range,
we go on to develop the  more applicable simple one dimensional model in section \ref{Simple}. This is later validated  it by making additional  detailed
comparisons  between the two approaches.
In the one dimensional model one can calculate the evolution of the disk surface density in the first instance assuming
the radius of the assumed circular orbit is a fixed parameter. One can  then use this solution to determine 
how the orbit evolves. It turns out that when a deep gap is formed only the disk exterior to the perturber has to be considered
as is  explained in section \ref{evoutd}.
A relevant similarity solution for an  accretion disk of  infinite extent is given in \ref{infdisc}.
and the time scale for  orbital evolution induced through the action of disk torques  discussed in section \ref{orbtim}.
In addition  accretion rate onto the perturber is estimated in section \ref{est}.

We compare the results of the one dimensional model regarding gap formation and  migration  for larger mass  perturbers,
with two dimensional numerical simulations in section \ref{Sims}.
 The two approaches are again found to be in good agreement.

We go on to briefly discuss additional  features that should be considered in future work, 
such as  possible  effects  of a moderate to large imposed orbital eccentricity,  orbital evolution driven by
the emission of gravitational waves, a process that may dominate when the perturber
reaches the inner regions of the disk,  and the secular evolution of the direction of binary's angular momentum due
to interaction with a twisted accretion disk, in section  \ref{Disc}.
Finally we summarise our  results and conclusions in section \ref{sumconc}
 
\section{Basic  definitions and set-up } \label{s1}
We consider  a binary consisting  of a  primary
of  mass $M$ and a secondary of mass $M_p$ that is  embedded in an accretion disk.  
The binary orbit is taken to be coplanar with the disk mid-plane,  approximately  circular and with a sense of rotation that
is opposite to that of the disk gas.
 We suppose that that the primary is much more massive than the secondary such that  the mass ratio
$q=M_p/M \ll 1.$ Thus disk material interior to the orbit of the secondary  revolves
approximately in circles centred on the primary.

We determine the modification of the disk structure due to the
presence of the binary                
and the evolution of  the binary separation distance, $r_p,$
induced by torques exerted by the disk material.
In order to do this we employ  a simple numerical 
approach based on  an azimuthally averaged  and hence simplified  one dimensional  description of the accretion disk 
as well as  an additional  simplified  analytic treatment of the problem.    
We go on to   relate these to
two-dimensional numerical simulations of the disk interaction with the binary.
In the work presented below, we adopt  a cylindrical coordinate system $(r,\phi, z)$ with origin at the primary and with the $z$ axis directed 
perpendicular to the orbital plane. We assume that the disk  material and the binary orbit 
with increasing  and  decreasing  azimuthal angle and thus in a prograde and retrograde sense   respectively.
\section{A one dimensional model for the evolution of the disk and binary orbit}\label{1Dmodel}
We develop a simplified one dimensional model for the evolution of the binary  orbit taking into account  the
gravitational interaction with  the disk  (e.g. Lin \& Papaloizou 1986)  which itself undergoes viscous  evolution
due to angular momentum transport (e.g. Lynden-Bell \& Pringle 1974).

\subsection{The evolution of the accretion disk and its interaction with the binary orbit}\label{evorb}

In  order to find the torque $T$ due to the disk acting on the binary it is necessary
to determine how the presence of the perturbing body affects  the structure of the  disk.
Following the discussion of Lin \& Papaloizou (1986) applicable to the case when the binary and disk
rotate in the same sense, we assume that  the  gravitational interaction results in a locally induced angular momentum
transport to the disk. This in turn, together with internal angular momentum transport
induced by  the action of an effective turbulent viscosity, determines the evolution of the
disk  surface density $\Sigma.$

The evolution of $\Sigma$ can be obtained from consideration of the conservation of mass and angular momentum.
The former is expressed by the continuity equation which  can be written as
\begin{equation}
{\partial \Sigma \over \partial t}= {1\over 2\pi r}{\partial \dot M \over
\partial r}, \label{e1}
\end{equation}
where the mass flux through radius $r$ is given by
\begin{equation}
\dot M =-2\pi \Sigma rv_{r},
\end{equation}
with  $v_r$ being  the radial velocity of the disk material.

The conservation of
the $z$-component of angular momentum is expressed by the equation
\begin{equation}
 r^2\Omega {\partial  \Sigma \over \partial t} ={1\over 2\pi r}
{\partial  \dot L  \over \partial r} + \Sigma  \dot J, \label{e2}
\end{equation}
where  the angular momentum flux through radius $r$ is given by
\begin{equation}
\dot L = 2\pi\left (\nu r^{3}\Sigma {d\Omega \over dr} - \Sigma r^3 \Omega v_{r}\right),
\end{equation}
with   $\nu $ being the  kinematic viscosity of the disk material,
$\Omega(r) =  \sqrt {GM/r^3}$
is its  angular velocity which is assumed to be Keplerian and
$G$ is the gravitational constant.
 The torque exerted per unit mass by the  perturber on the
disk is $\dot J.$
Thus the total torque exerted on the disk by the perturber is given by
\begin{equation}
T=2\pi \int rdr \Sigma \dot J.
\label{en3}
\end{equation}
 Note that unlike the prograde case there are no Lindblad resonances in the retrograde circular case, and, accordingly,
their contribution is not included in (\ref{en3}). However, they can operate when a retrograde binary has some
eccentricity. This effect is briefly discussed in section \ref{eccorb} below. 
\subsection{The evolution of the binary orbit}\label{s1.2}

By Newton's third law, the  total torque exerted on the binary will be $-T.$
Provided that it  remains  approximately circular,
angular momentum conservation  for the orbit  determines the evolution  of $r_p$ through
\begin{equation}
{\dot L_b}=-T, \quad {\rm with} \quad L_b=-qM\sqrt{GMr_p},
\label{en1}
\end{equation}
being the orbital angular momentum, which has a negative sign on account of the orbit being retrograde.
 From equation (\ref{en1}) we find
\begin{equation}
{\dot r_p}=-{r_p\over t_{ev}} \quad t_{ev}= {|L_b|\over 2|T|}.
\label{en2}
\end{equation}
which  defines a characteristic timescale, $t_{ev},$  for  the orbital evolution.

Note that for the configuration considered here, the  binary is expected to  transfer
retrograde angular momentum to the disk causing it to slowly spiral inwards.
Thus we expect $\dot J \le 0$ and  $T \le 0.$
As a consequence the disk  will gain retrograde, or equivalently  lose  prograde,  angular momentum.
Note that we expect  $\dot J \le 0$   everywhere.  Thus   unlike the case for which
the binary is prograde,  $\dot J $ does not change sign on crossing the orbit.

We now go consider two regimes of perturber disk interaction. The first, appropriate for small mass ratios,
is when the disk surface density is only slightly modified
such that the interaction can be regarded as linear. It  corresponds to the Type I migration regime.
The second, appropriate for large enough mass ratios, is when the interaction is nonlinear   such that a gap is formed in the disk. 
In this case, corresponding to Type II migration, migration rates become  significantly reduced relative to the Type I regime.

\section{Small mass ratios and type I migration}\label{s2}
In this case we assume that the perturber has a small enough mass such that the  disk  surface density remains unchanged  by  the embedded perturber,
at least  on long enough time scales such that significant migration can occur.
In this regime,  the torque  exerted on the perturber  through interaction with the disk occurs through the excitation, transport and dissipation
of  density waves. 
In order to find the resulting torque,  $T_{wave},$   exerted on the binary  we perform a linear calculation
of the disk response.  This approach is standard  when considering the type I  migration  regime in the prograde case
(see e.g. Baruteau et al. 2014).
 
We begin by  performing  a Fourier expansion of the gravitational
potential of the perturber, $\psi_p,$ in the form
\begin{eqnarray}
\psi_p& =& {\cal{R}}e\sum_{m=0}^{\infty} W_m\exp[{\rm i}m(\varphi +\omega t)]\quad {\rm with} \nonumber \\ 
 W_m &= &-\frac{GM_p}{\pi}\int_{0}^{2\pi}\frac{\cos m\varphi d\varphi}{\sqrt{r_p^2+r^2+\Delta_s^2 -2rr_p\cos\varphi}}\nonumber \\
 &\sim& -\frac{2GM_p}{\pi r_p}K_0\left( \frac{m\sqrt{\Delta^2+\Delta_s^2}}{r_p} \right) \hspace{2mm} {\rm for}\hspace{2mm} m > 0. \label{potexp}
\end{eqnarray}
Here $m$ is the azimuthal mode number we recall that $\Delta = |r-r_p|$, $\omega=\sqrt{{GM/ r^3_p}}$ is the binary orbital frequency
and $K_0$ denotes the modified  Bessel function of the second kind.
A  gravitational softening length, $\Delta_s,$ is included. This is regularly 
used to approximately account  for 3D effects (e.g. Baruteau et al. 2014).   
The representation of the Fourier coefficient through a Bessel function, as well as the neglect of the indirect term in the perturbing potential, 
should be accurate
 either for small $\Delta_{eff} = \sqrt{\Delta^2+\Delta_s^2}/r_p$  or   for $m\Delta_{eff}$ up to of order   unity
in the limit of large $m$. We  shall assume that  one of these conditions is appropriate   from now on.

For a barotropic equation of state, the disk response to the action of the perturbing potential (\ref{potexp})
can be obtained from   equations (45) and (46) of Lin \& Papaloizou (1993)
with forcing frequency applicable to a  secondary in a retrograde  circular  orbit.  These take  the form
\begin{eqnarray}
&&\frac{1}{\Sigma r}\frac{d \left(\Sigma r \xi_{r,m}\right)}{dr}- \frac{2\Omega \xi_{r,m}}{r(\Omega+\omega)}
=  \nonumber \\
&&K_m \left( \frac{1} {r^2(\Omega+\omega)^2}- \frac{1}{c^2}\right) +  \frac{W_m}{r^2(\Omega+\omega)^2}. \label{cr1}
\end{eqnarray}
and
\begin{eqnarray}
& &\frac{d K_m }{dr}+ \frac{2\Omega K_m }{r(\Omega+\omega)}
= \nonumber\\ 
& &\left[ m^2(\Omega+\omega)^2 - \Omega^2\right]\xi_{r,m} - \frac{d W_m }{dr} -  \frac{2\Omega W_m}{r(\Omega+\omega)}.\label{cr2}
\end{eqnarray}
Here $\xi_{r,m}$  and $K_m$ are the Fourier coefficients in the expansions analogous to (\ref{potexp})
for the radial component of the Lagrangian displacement and $W = \Sigma'c^2/\Sigma$ respectively, with
$\Sigma'$ being the induced surface density perturbation.
The  local sound speed is  $c$.
 \subsection*{Local approximation}
 As the  typical wavelength  of the density wave  response
is   $\sim c/(m\omega)\ll r$ 
even for $m=1$ we expect that they attain the  form  of  outgoing 
waves for $|r-r_p|/r_p \ll 1.$ Thus it is appropriate to 
look for local solutions for which outgoing wave boundary conditions are applied at radii
close to the orbital radius of the perturber. 

We  thus  assume $\Sigma$  and $c^2$ are  constant and replace $r$ and $\Omega$  where they  appear explicitly
 by $r_p$ and $\omega$ respectively.  
As $c^2 \ll r^2\omega^2,$ 
the  second term on the LHS and first term on the RHS of equation (\ref{cr1})
are  neglected. This then becomes
\begin{equation}
\frac{d \xi_{r,m}}{dr}
=- \frac{K_m}{c^2} +  \frac{W_m}{4r^2\omega^2} \label{cr11}
\end{equation}
and using the same  approximation scheme  (\ref{cr2}) becomes
\begin{equation}
\frac{d K_m }{dr}
=(4m^2-1) \omega^2\xi_{r,m}  -   \frac{d W_m }{dr} -  \frac{ W_m}{r}.\label{cr21}
\end{equation}

We now use (\ref{cr11})  to eliminate $K_m$ in (\ref{cr21}) noting
that in our approximation scheme it turns out that  the  second term on the RHS of (\ref{cr11})  may be neglected.
We thus obtain a  governing equation for   $\xi_{r,m}$  of the form
\begin{equation}
c^2 \frac{d ^2 \xi_{r,m} }{dr^2}
=- (4m^2-1) \omega^2\xi_{r,m}  +  \frac{d W_m }{dr} + \frac{ W_m}{r}.\label{cr3}
\end{equation}
This is seen to be an  equation for a forced  simple harmonic oscillator.
However, it is important to  note that, through its dependence 
on the gravitational potential of the  perturber,  the effective forcing  term 
involving $W$ varies rapidly in its  vicinity 
and cannot be assumed to be constant. 
It is convenient to  write the governing
equation in the compact form
 \begin{equation}
 \frac{d ^2 \xi_{r,m} }{dr^2}
=- k^2\xi_{r,m}   + S,     \label{cr4}
\end{equation}
where $k^2=(4m^2-1)\omega^2/c^2$ and

\noindent   $S=~(~dW_m/dr~+~W_m/r_p)~/~c^2~.$
\subsection*{Solution of the governing equation for the linear response}

Equation (\ref{cr4})  is solved subject to radiation  boundary conditions.
For convenience we adopt the convention that $\omega$ and $k$ are positive.
The free solutions of (\ref{cr4}) corresponding to inward and  outward propagating density
waves are then
\begin{eqnarray}
 \xi_{r,m}& =& C_1 \exp({\rm i}kx)\quad {\rm  and}\nonumber \\
\xi_{r,m} &= &C_2 \exp(-{\rm i}kx),
\end{eqnarray}
where $x=r-r_p$ and $C_1$ and $C_2$ are arbitrary constant amplitude factors  respectively
and real parts are taken to obtain  physical solutions here and below when needed.
The solution to the forced problem is determined such that it
 takes a multiple of  the above forms at large distances
(measured in terms of wavelengths) interior to and exterior to  the source, respectively.

The required solution is  found by standard methods to be given by
\begin{equation}
 \xi_{r,m}
=\frac{{\rm i}}{2k}\int^{\infty}_{-\infty} \exp(-{\rm i} k|x-x'|) S(x')dx' ,     \label{cr5}
\end{equation}
with
the constants $C_1$ and $C_2$ which   specify the amplitude of the inward and outward
propagating waves at large distances   being  readily seen to be given by
\begin{eqnarray}
C_1&=&\frac{{\rm i}}{2k}\ \int^{\infty}_{-\infty} \exp({-\rm i} kx') S(x')dx' \nonumber \hspace{3mm} {\rm and}\\
C_2&=&\frac{{\rm i}}{2k}\ \int^{\infty}_{-\infty} \exp({\rm i} kx') S(x')dx' .\label{coeffs}
\end{eqnarray}

\subsection*{Rate of angular momentum transport}
Each of the waves described above transfers angular momentum from the orbit  to the disk
at the locations where they eventually dissipate. In a low viscosity disk, dissipation
occurs  as a result of non linear steepening and shock formation.
Depending on their amplitudes, this may be some distance away from the location of the orbit.
As the orbit is retrograde with respect to the disk, the  transfer removes positive angular momentum
from the disk, potentially causing it to accrete onto the central object.
The rate of flow of angular momentum
through a circle of radius, $r,$ associated with either of the  waves  is
\begin{equation}
F_J= -{\cal {I}}m \left(\pi m c^2 r_p \Sigma \xi_{r,m} \frac{ d \xi^*_{r,m}}{dr}\right) ,
\end{equation}
where ${\cal{I}}m$ denotes that the imaginary part is to be taken.

\noindent We evaluate  this expression for ingoing and outgoing waves with a particular value of $m.$
These  both remove negative angular momentum from the  binary orbit  and 
ultimately transfer it to the disk,
accordingly giving additive contributions  to  the
torque acting on  the disk which we write as $-T_{wave}.$
The corresponding  torque acting on the perturber will  then be  $T_{wave},$ which is positive.  
On account of it being retrograde,  this acts so as to make the perturber  spiral inwards.
For  a particular value of $m,$   we thus obtain
\begin{equation}
T_{wave} =  \pi m c^2kr_p  \Sigma (|C_1|^2+ |C_2|^2).\label{wavet}
\end{equation}

This  quantity is directly related to the Fourier transform of the source, $S(x),$
through equation (\ref{coeffs}).
A useful result  that enables  its evaluation 
is provided by the standard integral
\begin{eqnarray}
&& \int^{\infty}_{-\infty} \exp({\rm i}kx) K_0\left( \frac{m\sqrt{x^2+\Delta_s^2}}{r_p} \right)dx
= \nonumber \\
&& \frac{\pi}{\sqrt{k^2+m^2/r_p^2}}\exp \left (-\Delta_s {\sqrt{k^2+m^2/r_p^2}}\right).\label{FT}
\end{eqnarray}
Making use of  (\ref{coeffs}) and  (\ref{FT}), from equation (\ref{wavet})    we find that
\begin{equation}
T_{wave} =\frac{2\pi\Sigma G^2M_p^2 (1+k^2r_p^{2})}
{m r_p kc^2  (1+ k^2r_p^2/m^2)} \exp \left (-2\Delta_s {\sqrt{k^2+m^2/r_p^{2}}}\right) .\label{Jdot}
\end{equation}
As  $k >> m/r_p$ for all $m \ge 1,$ we simplify   (\ref{Jdot}) to read
\begin{equation}
T_{wave} =  \frac{2\pi m \Sigma r_p^4}{ \sqrt{4m^2-1}} q^2\delta^{-1}\omega^2  
\exp \left (-2\Delta_s \sqrt{4m^2-1}/H\right) \label{Jdotsimp}
\end{equation}
 and we recall  that $q=M_p/M$ and $\delta =H/r.$
This shows that when $\Delta_{s} \sim H,$ as is expected to be appropriate
for approximately accounting for the finite disk thickness and as adopted
 in  many of our numerical simulations,
the dominant contributions come from the smallest $m.$
This is unlike the prograde case for which the dominant contributions come from  $m \sim r_p/H.$
Taking into account only $m=1$, we  calculate a migration time for the perturber  using (\ref{en2})
with $|T|=|T_{wave}|,$ thus obtaining
\begin{equation}
\frac{-\dot r_p }{r_p} = \frac{1}{t_{ev}} =  \frac{4\pi \Sigma r_p^2}{ \sqrt{3}}\left(\frac{r_p}{H}\right) \frac{M_p}{M^2}\omega
 \exp \left (-2\sqrt{3}\Delta_s/H\right) .\label{tmig}
\end{equation}
Comparing this expression  with the corresponding one given by  Tanaka et al. (2002)
for the prograde case and a uniform surface density, we find that
$t_{mig}$ is slower by a factor
\begin{eqnarray}
\frac {2.7\sqrt{3} }{4\pi}\left( \frac{r_p}{H}\right)\exp \left (2\sqrt{3}\Delta_s/H\right) \sim 59 \nonumber
\end{eqnarray}
for the  parameters  of two dimensional simulations we performed to test (\ref{tmig}).

\subsection{Total torques obtained by summing over $m$}
When a sum of the  contributions from  all  values of $m$  is performed, the
analytic theory indicates that total torque  should be $\propto \Delta^{-1}_s$  as the softening length tends to zero.
However, our derivation of the expression (\ref{Jdotsimp}) relies on the assumption that the linear theory
is valid. Clearly, this falls  when a typical  inverse wavenumber  associated with the disk response,  $k^{-1} \sim \delta r_p/(2m),$
becomes comparable to or less than  the accretion radius $r_{a} \sim q r_p/2$ (this defines the impact parameter  for significant
scattering if the response were purely ballistic).
Using the equality $r_{a}=k^{-1} $ to define  a maximal
cutoff value of $m,$  $m_{max},$ we obtain
$m_{max} \sim \delta /q.$
Summing  contributions from different values of $m$  given by (\ref{Jdotsimp}) up to this cut off, 
we obtain  an estimate of the total
torque corresponding to the low mass retrograde case as
\begin{equation}
T_{wave}^{r}=\sum^{m_{max}}_{m=1}T_{wave} \sim \pi q \Sigma \omega^2 r_p^4.
\label{en14}
\end{equation}

\subsection{Type I migration rates}\label{Tir}

We remark that  (\ref{en14}) gives the migration torque  with the  cut off scale  assumed to
be governed by the  accretion radius. However, if the accretion radius is less than the scale height
the cut off is more likely to be determined by the latter.  To deal with  that  case we  replace $r_{a}$  by $0.3\pi H$
in the above  determination of $m_{max}.$
To take account of both possibilities,  we multiply the torque given by   (\ref{en14}) by a factor $f,$
where $f = min(1, 5q/(3\pi \delta)).$ Thus
\begin{equation}
T_{wave}^{r}=\pi f q \Sigma \omega^2 r_p^4.
\label{en15f}
\end{equation}
It is instructive to compare (\ref{en15f}) with the   corresponding expression for the prograde case,
\begin{equation}
T^p_{wave}\sim {q^2\over \delta^2}
 \Sigma \omega^2 r_p^4.
\end{equation}
From (\ref{en15f}) we see that the ratio
\begin{equation}
T_{wave}^{r}/T_{wave}^{p}\sim {\pi f \delta^2 \over q}.
\label{en15a}
\end{equation}
Thus, low mass objects immersed in the disk and rotating in different directions
could drift with different radial velocities determined by the disk half-thickness and
their  mass ratios.  In principal, this could lead to some interesting consequences, as e.g. a possibility of collision between
the objects since, in this situation, for comparable mass ratios,
the  radial separation between them  will in general decrease relatively rapidly  with time.

\section{Numerical simulations of embedded perturbers}\label{Esim}
We have  performed numerical simulations  using
the 2D code NIRVANA (see e.g. Podlewska-Gaca et. al. 2012  and references therein)
and also the ROe solver for Disk Embedded Objects (RODEO) method
(see e.g. Paardekooper \& Papaloizou 2009).
In the context of results presented here, these independent methods were found to give almost identical results.

In this section we  consider  numerical simulations for which the perturber remained embedded  without forming a significant gap.
The initial surface density was specified to be $\propto r^{-1/2}$
and scaled so that the total mass interior to the initial orbital radius of the perturber was $10^{-3}$ in units of the dominant central mass.
The perturber was initiated on a retrograde circular orbit of radius $r_0$ which is taken to be the simulation  unit of length.
For simulation unit of time we take the orbital period of a circular orbit with this radius.
We adopted   $H/r =0.05$
and the kinematic viscosity
was taken to be $\nu = 10^{-5}r_0^2\Omega(r_0).$ 
 The computational domain
was taken to be $(0.2r_0, 5r_0).$ 
The outer radial boundary was taken to be rigid and the inner radial
boundary open.
 Gravitational  softening lengths ranging from  $\Delta_s = 0.6H,$ denoted as standard softening,
to  $\Delta_s = 0.1H,$ denoted as small softening were considered.
Grid resolutions with $N_R=350$ and $N_{\phi}=400$ equally spaced grid points in respectively
the radial and azimuthal directions
were typically  adopted
for cases with standard softening.
For smaller softening lengths 
the resolution was increased to $N_R=700$ and $N_{\phi}=1200.$
For simulations presented in this section, there was no accretion onto the perturber.

\begin{figure}
\resizebox{\hsize}{!}{\includegraphics[angle=270]{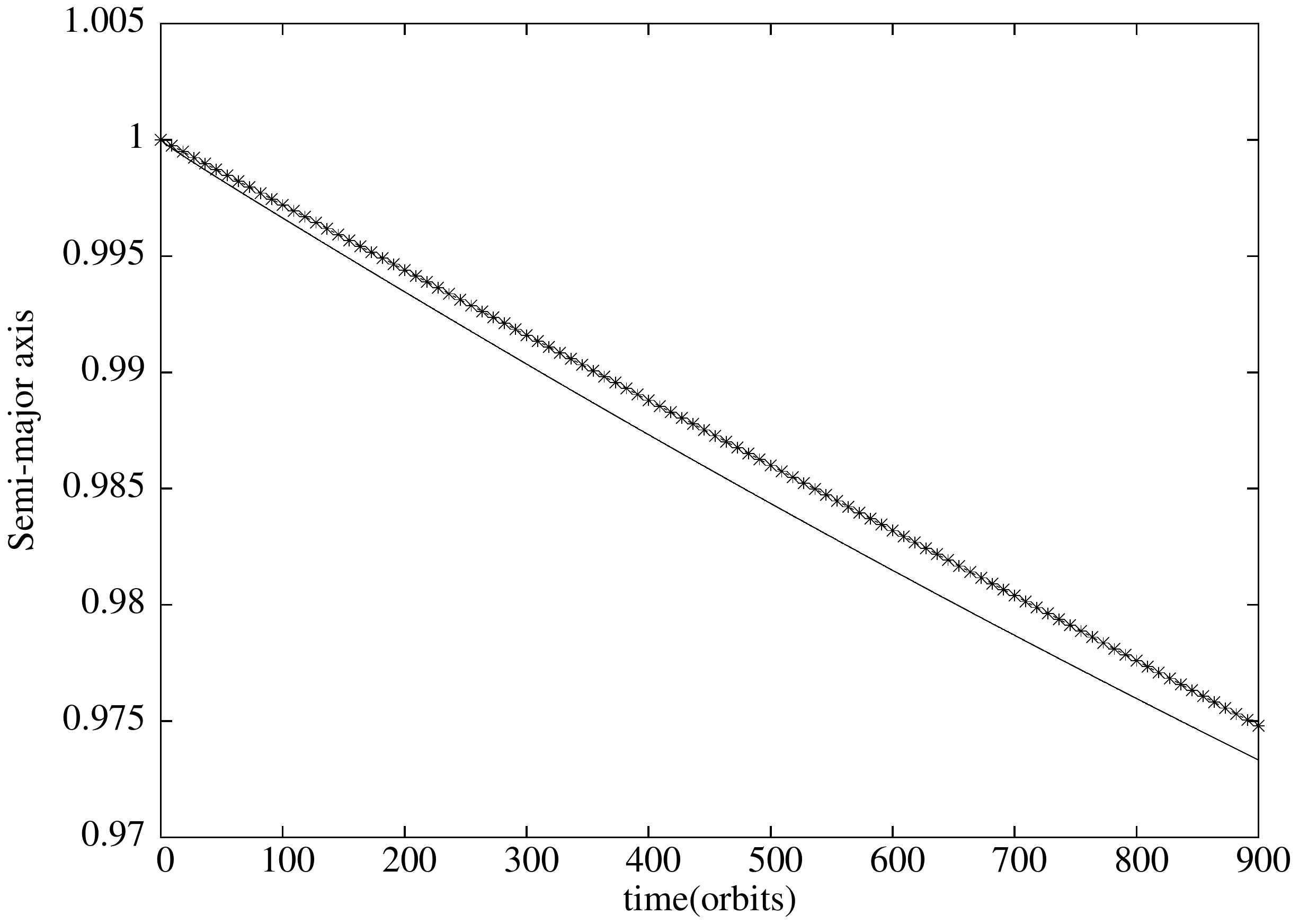}}
\caption{ Semi-major axis, in units of the initial orbital radius as a function of time  orbits for $q=0.001.$ 
The solid curve  give the results from numerical simulation and the curve  with imposed crosses  is obtained   using  equation(\ref{en2})
 with   contributions to the total torque for $m \le 100$  obtained from equation (\ref{Jdotsimp}).
However, the main contribution comes from $m=1$ in this case (see  Section \ref{typeirates} below).
The numerical simulation was performed with no accretion onto the companion.
} \label{Migratione-3.eps}
\end{figure}

The results of a  numerical simulation that was performed  for $q=0.001$  for standard
softening and  with no accretion onto the perturber are illustrated in Fig.\ref{Migratione-3.eps}.
 The semi-major axis, in units of the initial orbital radius is plotted as a function of time.
  In addition  the evolution expected from  the analytic theory, obtained from  
equation (\ref{en2}) with $|T|$ evaluated  by summing  the values of $T_{wave}$ calculated using  (\ref{Jdotsimp}) 
 for  $m\le 100.$   

\begin{figure*}
\centering
\includegraphics[width=17cm]{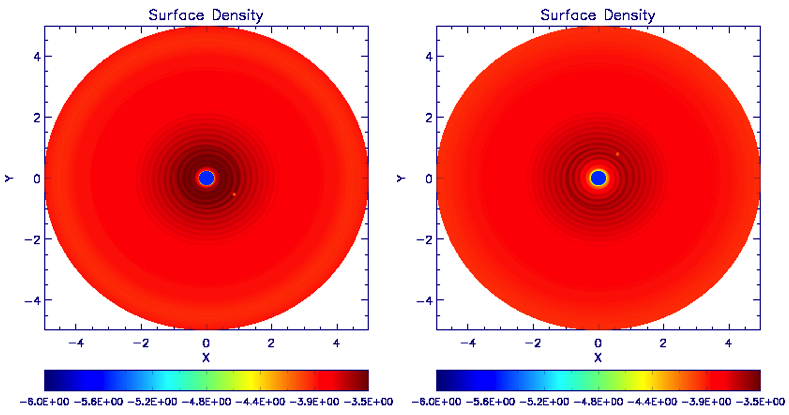}
\caption{  $\log\Sigma$  contours for $q=0.001$ with softening length $0.6H$ after $100$ orbits (left panel) and after $840$ orbits (right panel). In these simulations the companion, its position in each case being at the centre of  the small superposed red circle,  located on the line at an angle of $\sim 315^{\circ}$ to the $x$ axis (left panel) and  at an angle of 
 $\sim 45^{\circ}$ to the $x$ axis (right panel) was not allowed to accrete. Short wavelength density waves are visible on both sides of the quasi-circular orbit. The relative density changes are similar in both these plots. However, the density in the interior regions of the disk decreases at later times on account of accretion through the inner boundary.} \label{1e-3100P}
\end{figure*}

There is seen to be very good agreement between the two with the total amount of radial migration
differing by less than $10\%$ over $900$ orbits.
The form of the surface density after $100$ and $840$  orbits is illustrated in  Fig.\ref{1e-3100P}.
Apart from the central regions where the surface density   becomes small on account of the outflow boundary condition,
the surface density remains relatively unperturbed with low amplitude density waves being apparent in the vicinity of the orbit.

\subsection{The dependence  on  the mass ratio and  softening length}\label{typeirates}

\begin{figure}
\resizebox{\hsize}{!}{\includegraphics{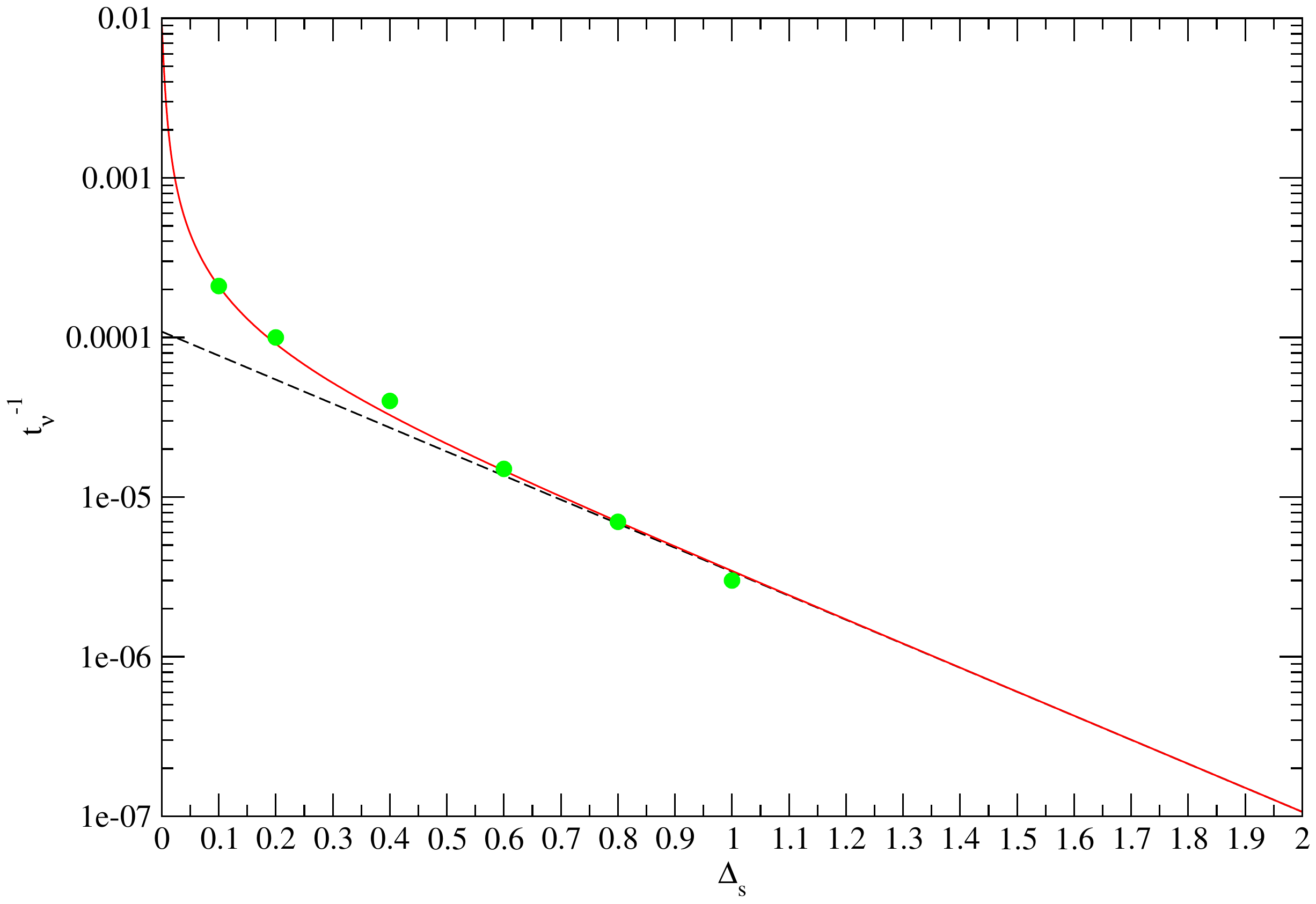}}
\caption{The inverse of the evolution time scale defined according to equation
(\ref{tmig}), as a function of the dimensionless softening length $\Delta_s/H$. It is expressed
in  units of   $P_0^{-1}$, where $P_0$ is the  initial orbital period of the perturber.
The dashed line is calculated  adopting  equation (\ref{Jdotsimp}) with only the contribution with  $m=1$ included.
The solid line is calculated including  contributions from all  $m\le 100.$
The  circles indicate  the results of numerical simulations.
} \label{SJ}
\end{figure}

 A comparison of orbital evolution rates
calculated analytically  as indicated above
with those obtained from our numerical simulations for different mass ratios with standard softening and no accretion
are presented  in Table \ref{table1}.  In addition Fig. \ref{SJ}  indicates  the contributions from $m=1$ and all $m < 100$ to the analytic evolution rates.
There is good agreement between the numerical and analytic results given in Table \ref{table1}  which differ by not more than $25\%,$
with the maximum deviation occurring for the smallest mass ratio case.

\begin{table}[h]
\begin{center}
\begin{tabular}{ l | c | r }
q&  $(dr_p/dt)_{n}$ &   $(dr_p/dt)_{a}$\\

0.0001&  -2.1e-6&-2.8e-6\\
0.0005 & -1.6e-5& -1.4e-5\\
0.001  & -3.4e-5& -2.8e-5\\
0.003  & -7.9e-5& -8.3e-5\\
\end{tabular}
\end{center}
\caption{The Table shows the evolution rates calculated numerically,  $(dr_p/dt)_{n}$, and
analytically as indicated above,  $(dr_p/dt)_{a}$, for different values of the mass ratio $q$. The evolution
rates,  time averaged over several orbits and evaluated at early times,
are in units $r_0 P_0^{-1}$, where $P_0$ is the  initial orbital period of the perturber. The softening length
for these calculations was taken to be $0.6H.$}
\label{table1}
\end{table}

The dependence on the softening length was also investigated for the case with $q=0.005,$
and the results of a comparison between numerical and analytic migration rates are shown in Fig. \ref{SJ}.
One can see that in this case  there is a very good agreement between the analytical and numerical results
down to softening lengths $\sim 0.1H.$ 

\section{Gap formation and migration for  higher mass perturbers}\label{Nonlin}
We now consider perturbers  massive enough  to make  the interaction between the disk and perturber  non linear.
We first consider simplified semi-analytic models for calculating gap profiles and 
migration rates,  subsequently comparing results with direct numerical simulations.

\subsection{ A simplified  description of the interaction between the perturber and disk}\label{simpdesc}
In general $\dot J$  can be  determined by considering the disk  response
to the perturber as disk gas streams by.
When dissipation is efficient enough to prevent angular momentum being carried away by waves,
the process can be modelled as a
 direct transfer of 
angular momentum to the gas particles when they are  scattered  by
the perturber while  undergoing  a  close approach to it (see Lin  $\&$ Papaloizou 1979).
When dissipation is ineffective, the transferred angular momentum is carried away by waves
which may dissipate in some other location making the process non local
(see e.g. Papaloizou \& Lin 1984).
When the disk viscosity is small, waves are expected to be excited when the perturbation is linear.
On the other hand when the perturbation is non linear, shocks are likely to occur
resulting in the local transfer of angular momentum. 

  For the regions of the disk where dissipation is effective locally, we adopt the
impulse approximation employed by  Lin \& Papaloizou (1979) 
( see also Papaloizou \& Terquem 2006) in the prograde case.
However, 
here we modify the analysis to allow for the fact that  the perturber moves in the
direction opposite to that of the gas.
Then
 we find that approximately
\begin{eqnarray}
\dot J& = & -{1\over 4\pi}q^{2} (\Omega r)^{2}{r^{2}\over
\Delta^2 +\Delta_s^2} \quad {\rm  when} \quad |\Delta | < r_c \quad  {\rm and}  \nonumber \\
 \dot J &= & 0 \quad {\rm when } \quad |\Delta | > r_c. \label{e3}
\end{eqnarray}
Here $\Delta = r-r_p,$  and we have introduced a gravitational softening length,
$\Delta_s,$  which  prevents a divergence when  $\Delta \rightarrow 0.$

The cut off radius 
$r_c$ gives the distance inside which the perturbation is non linear
and the impulse approximation can be employed. At greater distances we assume
that the angular momentum transferred is carried away by waves and so does not affect the disk locally.
 We estimate that
$r_c =min
(r_{h}, r_{s})$, where $r_h \approx (q/3)^{1/3}r_p$ is the Hill radius.
The 'sound' radius, $r_s,$  is defined through  $r_s = (q/ \delta ) r_p,$ 
 where $\delta = H/r$ with  $H$ being  the disk  semi-thickness. This  radius
is defined as the impact parameter such that
 the radial velocity component of a gas
particle induced by the scattering event is  equal to  the sound
speed. 
When the sound radius  exceeds the Hill radius, the latter  is taken
to be the cut off distance.

Note that the angular momentum carried away by waves does not affect the disk locally
but should be taken into account when considering the evolution of the orbit.
However, this is only a small effect once the interaction becomes significantly non linear
as linearity  then only applies at large scattering impact parameters for which the interaction is weak. 
Thus in that regime it may be neglected. 

 We here remark that the linear estimate (\ref{en14}) can be obtained from the above  arguments based on treating the disk  response
to the perturber using the impulse approximation.  To obtain it equations
(\ref{en3}) and (\ref{e3})  are used under the assumption that the latter is approximately valid even  when $\Delta \sim r_a .$
 Approximating
 (\ref{en3}), taking into account the disk on both sides of the perturber,  as
\begin{equation}
T=  4\pi\Sigma r_p\int^{\infty}_{ r_{min}}  \dot J d\Delta
\end{equation}
 and making use of  (\ref{e3}) with $\Delta_s=0,$  we recover
the expression (\ref{en14}) when  the inner  cut off radius, $r_{min} \ll r_c$  is set equal to $2r_a/\pi.$

\subsection{Evolution of the disk surface density}\label{evdiscsig}

Multiplying equation (\ref{e1}) by $r^2\Omega $ and subtracting the result 
from (\ref{e2}) we  obtain an expression for  $v_r$ in the form
\begin{equation}
v_r={2\dot J\over \Omega r}-{3\over \Omega r^2 \Sigma}{\partial \over
\partial r} (\nu \Omega r^2 \Sigma ), 
\label{en6}
\end{equation}
and, accordingly,
\begin{eqnarray}
\dot L& =& 2\pi r^2\left (3\Omega r {\partial \over
\partial r} (\nu \Sigma ) - 2\Sigma \dot J\right ), \quad {\rm and} \nonumber \\ 
\dot M &=& {2\pi \over \Omega} \left ({3\over r}
{\partial \over \partial r} (\nu \Omega r^2 \Sigma ) - 2\Sigma \dot J\right).
\label{en7}
\end{eqnarray}
Substituting equation  (\ref{en6}) into  equation (\ref{e1}) and
making use of  (\ref{e3}) we  obtain a single equation for the
evolution of the  surface density in the form
\begin{equation}
{\partial \Sigma \over \partial t}  =  {1\over r} {\partial \over
\partial r}\left (  {3\over \Omega r} {\partial \over
\partial r}(\nu \Omega r^2 \Sigma ) +        
{q^2\Theta (D) \Sigma \Omega r^4\over 2\pi
(\Delta^2 +\Delta_s^2)} 
\right),\label{e4}
\end{equation}
where $\Theta (D) $ is the step function and $D=r_c-|\Delta|$.

It is convenient to introduce dimensionless variables $\tilde r = r/r_0$ and
$\tilde \Sigma = \Sigma /\Sigma_0$, where $r_0$ and $\Sigma_0$ are the  initial orbital radius
of the perturber 
and surface density  at  its initial location, 
respectively. In addition  we   use $h=\sqrt{\tilde r}$  as  spatial coordinate.
We also assume power law dependences of 
 $\nu$ on $r$ and $\Sigma $ through the relation
\begin{equation}
\nu =\nu_* \Sigma^a r^b=\nu_0 {\tilde \Sigma}^a {\tilde r}^b=\nu_0{\tilde \Sigma}^a h^{2b}, \label{e5}
\end{equation}
where $\nu_*, $ $\nu_0,$  $a$ and $b$ are constants. 
Then  equation (\ref{e4}) takes the form
\begin{eqnarray}
{\partial \tilde \Sigma \over \partial \tau} &  =  & {1\over h^3}
{\partial F \over \partial h} , \quad {\rm with}\nonumber \\ 
F&=& {\partial \over
\partial h}(h^{1+2b} {\tilde \Sigma}^{1+a} )+\beta \Theta (D) {h^5
\tilde \Sigma \over {\tilde \Delta}^2 +{\tilde \Delta_s }^2}
 ,\label{e6}
\end{eqnarray}
and
\begin{equation}
\tau=t/t_{\nu}, \quad  {\rm with }\quad t_{\nu}={4r_0^2\over 3\nu_0}. \label{e7}
\end{equation}
Here  $t_{\nu}$ is a characteristic time scale of viscous evolution of the
disk  at $r\sim r_0$, $\beta =q^2\Omega_0r_0^2/
(3\pi \nu_0)$, $\Omega_0=\Omega (r_0)$, $\tilde \Delta =\Delta/r_0$ and  $\tilde
b =b/r_0.$

From equations (\ref{en7}) and (\ref{e6}) it  then  follows that the  mass and angular momentum
fluxes can be expressed in the form
\begin{equation}
\dot M =\dot M_* F, \quad \dot L =\dot L_* h(F-h^{2b}{\tilde \Sigma}^{1+a}),
\label{en8}
\end{equation}
where $\dot M_*=3\pi \Sigma_0 \nu_0 $ and $\dot L_* =\dot M_* \Omega_0r_0^2$.

For our estimates below we use the standard representation of $\nu_0$ through the Shakura-Sunyaev 
parameter $\alpha $ ( Shakura (1973), Shakura
$\&$ Sunyaev (1973))as 
\begin{equation}
\nu_0=\alpha \delta^2 \Omega_0 r_0^2.
\label{en8add1}
\end{equation}
The coefficient $\nu_*$  defined through (\ref{e5}) can then be expressed in terms of quantities characterising opacity law in the disk and
$\alpha$, for an explicit expression see Lyubarskiy $\&$ Shakura (1987)  and IPP. 
From equation (\ref{en8add1}) we  obtain 
\begin{equation}
t_{\nu}\approx \alpha^{-1}\delta^{-2}\Omega_{0}^{-1}\quad \beta \approx 0.1 {q^{2}\over \alpha \delta^2} 
\label{en8add2}
\end{equation}

\subsection{Initial and boundary conditions for a disk structured by a perturber}\label{initbdry}

We assume that the disk is in a steady state at  time $t=0$ 
when the binary is introduced.  In the steady state
 the dependence of $\tilde \Sigma $ on $h$ follows from equations (\ref{e6}) with time
 derivatives and $q$  set to zero.  Thus
$d F / dh=0$, and therefore
\begin{equation}
\Sigma =\Sigma_0 h^{-{1+2b\over 1+a}}(F_0(h-1)+1)^{{1\over 1+a}},
\label{en9}
\end{equation}
and 
\begin{equation}
\dot M=\dot M_* F_0, \quad \dot L=\dot L_* (F_0-1),
\label{en10}
\end{equation}
where $F_0$ is a constant of integration  and we  have ensured that $\Sigma (h=1)= \Sigma_0$  as required by definition.
Of special interest are the  cases with  $F_0=1$ and $F_0=0$. These correspond to  cases with
zero angular  momentum flux and zero mass flux through the disk, respectively.  
The former case corresponds 
to a disk of formally infinite extent with constant  mass flux  equal to $\dot M_*.$ 
 Recalling  that the  angular momentum flux in a stationary disk
is determined by an inner boundary condition and is typically small,
we can consider the case $F_0=0$ as  appropriate
for astrophysical systems of interest such as a disk interacting with  a binary black hole 
since we expect  the secondary to  be immersed in the disk at radii 
much larger than  its inner boundary  radius
\footnote{For example,  for a  black hole and a 'standard' accretion disk we have 
$\dot L = \dot M \sqrt{GMr_{ms}}$, where $r_{ms}$ is the radius of the marginally stable orbit. Clearly,
it is much smaller than $\dot L_*$ when $r_0  \gg r_{ms}$.}. 

The  case  with $F_0=0$ can approximately describe a circumbinary disk around a massive binary rotating in 
the same sense as the disk  gas (e.g. Ivanov, Papaloizou $\&$ Polnarev 1999). 
Although there is no direct relation to situations  
considered in this Paper,  we use   $\tilde \Sigma $  distributions for such models in several numerical runs to
test different initial conditions. 

It is instructive to express the steady state surface density and the angular momentum flux
in terms of the mass flux, the quantity $h_*=(F_0 - 1) /F_0$
and the viscosity coefficient $\nu_{*}$ defined through  (\ref{e5}). We obtain
\begin{eqnarray}
\Sigma &=& \left({\dot M\over 3\pi \nu_*r_0^b}\right)^{1/( 1+a)} h^{-(1+2b)/( 1+a)}(h-h_*)^{1/( 1+a)}
 \hspace{2mm}  {\rm  with }\nonumber \\
 \dot L&=&\dot M \Omega_0r_0^2 h_*.
\label{en10a}
\end{eqnarray}
From  (\ref{en10a}) it is clear that $h = h_*$   corresponds to the  inner edge 
of the disk, where the surface density drops to zero.

Because we cannot perform two dimensional numerical simulations of accretion disks of arbitrary  radial extent,
we develop an approximate theory of   the evolution of the disk and  orbit 
that is  valid both for disks of  finite and infinite extent. 
This  will be tested against numerical  simulations for  the case of 
 disks of finite extent.

\subsection{Conditions for gap formation}\label{Gapf}
The action of the impulsive torque per unit mass  exerted on the disk
by the perturber  given by (\ref{e3})  is to cause gas elements to lose angular momentum as they encounter 
and are scattered by the perturber.  This causes them to move to smaller radii, enhancing any  inward  drift
resulting from viscous evolution. 
Since the disk gas in the vicinity of the orbit is supplied from
the  outer regions  of the disk at a rate determined  by viscous evolution and the presence of the perturber 
increases the magnitude  of the  radial velocity, $v_r,$ 
that is directed inwards, a surface density depression  must form close to  the 
perturber orbit 
in order that  the continuity equation be satisfied.
 We  hereafter describe this depression  as 'a gap' but emphasise that
the way  the gap is formed  differs from that   applicable to the  well known case when
the perturber is in a   prograde orbit (see eg. Papaloizou \& Terquem 2006).

As indicated above,  the impulsive torque is efficient only when the distance from the perturber, $\Delta $ is
smaller than both the Hill radius $r_h\approx (q/3)^{1/3}r_p$  and 
'the sound radius',
$r_s  \sim (q /\delta)r_p.$ 
When the mass ratio is very small, and, accordingly,  impulsive interactions are
not effective, the angular momentum transferred between the perturber and disk is transported away by waves
 and a pronounced gap in the disk is not produced.
 
  We make a simple estimate for when this should occur
by requiring that $r_s$ should be smaller than $\pi\delta r_p/2,$ which is half the longest  wavelength 
associated with  density waves launched by the perturber which occurs for $m=1$ (see above). This gives
 $r_s <   \pi\delta r_p/2 $
 from which we obtain 
\begin{equation}
q < q_1, \quad q_1 = {\pi \over 2}\delta^{2}\approx 1.57\delta^{2}.
\label{en11a}
\end{equation}

We emphasise that on account of it being obtained
from simple estimates,  this criterion is uncertain to within a numerical factor of order unity.
Nonetheless we found  that our numerical results are in agreement with it. We adopted  $\delta=0.05$ in our numerical
calculations. Then we obtain  $q_1\approx 5\cdot 10^{-3}.$ We  observed  gap   formation in the case of the somewhat
larger $q=0.01$ and did not observe it for $q=0.001.$   

\subsubsection{ Secondary mass larger than the local disk mass}
In the opposite limiting case of a large mass ratio  such that
 the secondary mass $M_p$ is much larger than a characteristic  disk mass within its orbital radius
$M_{d}\equiv \dot M_{*}t_{\nu}=4\pi \Sigma_0 r_0^2$, the local viscous evolution  time scale of the disk is expected to be
much smaller than the evolution time scale of the binary orbit. In this case one may find the time 
scale for evolution of the orbit  in two steps.
 At first, one can determine the  modification of the disk structure induced by  the binary assuming that its
separation distance $r_p$ is fixed and then calculate the interaction torque $T.$
One can use this in  equation (\ref{en2}) 
to find the evolution time scale.  From the condition $M_p > M_{d}$ we obtain
\begin{equation}
q > q_2, \quad q_2 = {4\pi \Sigma_0 r_0^2\over M}.
\label{en12}
\end{equation}
\subsubsection{Issue of gravitational stability}
Note that the ratio of the local disk mass, $M_d,$  to the mass of the primary should be of the order of or smaller than 
$\delta $ for the disk to be  gravitationally stable according to the Toomre stability criterion. It is, therefore, sufficient 
to have 
\begin{equation}
q > q_{crit}=\delta
\label{en13}
\end{equation}
for $q$ to be larger than both $q_1$ and $q_2$. 
Note that this case is analogous 
to what is considered in  Ivanov et al. (1999)  for systems with  prograde rotation. For
supermassive black hole masses appropriate to galactic centres inequality (\ref{en13}) typically holds.  

\section{ Simple 1D modelling  of the surface density profile for the case of relatively large mass ratio}\label{s3}
In this Section we assume that although the mass ratio $q\ll 1$ the conditions
 $q > q_1$ and  (\ref{en12}) are  both fulfilled, and, accordingly, the ratio
of the perturber mass to a characteristic mass of the disk is
large.  As we discussed above, in this situation  the
characteristic time for  evolution of the perturber's orbit is much larger
than that of the disk, so that in order to model the evolution
in a simple way, we can calculate quantities
characterising the evolution of the system iteratively, assuming 
at first that the  perturber's orbital distance is fixed and analysing 
properties of the disk, and then calculating the evolution rate of 
the binary.

\subsection{Structure of the gap around perturber's orbit obtained  by solving the diffusion
 equation incorporating the effect of torques due to the perturber }\label{gapstructure}

As we have seen,  a depression in the
profile of the surface density called a gap is formed in the neighbourhood 
of  the perturber's orbit. Gravitational interaction 
with the perturber removes angular momentum from disk gas elements as they stream past it,
 thus increasing their radial drift velocity in the vicinity of
the orbit. 
The formation of the gap can be described by  equation 
(\ref{e6}) which incorporates the effects of viscosity and torques due to the perturber. 
 It may  be shown that after some relatively short period of
time the solution to (\ref{e6}) in the neighbourhood of $r\sim r_p$ becomes quasi-stationary.  This implies  that in order to find the form 
of $\tilde \Sigma ,$
we can assume that the dimensionless mass flux, $F,$ defined there,  does not depend on the radial coordinate
$h$, and, accordingly, obtain an  equation
for $\tilde \Sigma $ of the form
\begin{equation}
\frac {d}{dx}  \left({\tilde \Sigma}^{1+a} \right)+  \beta {\tilde \Sigma \over 4{ x}^2 +{\tilde \Delta_s}^2} = F, \label{e8}
\end{equation}
where $x=h-1$, the ratio $r_c/r_0$ is assumed to be small and we
also consider the region, where $\tilde \Delta \le r_c/r_0$, and, accordingly,
the coordinate $x$ should be such that $x_{-}~\le~x~ \le~x_{+}$, where $x_{\pm}=\pm  r_c/( 2r_{0})$. 

\noindent Note that the condition (\ref{en13}) together with equation (\ref{en8add2}) imply that for $\alpha < 0.1$, 
$\beta $ should be of the order of 
or larger than unity,  and in the limit  that $q \gg q_{crit}$ we  accordingly have $\beta \gg 1$.

\noindent Since a detailed shape of $\tilde \Sigma $ close to the orbit does not
influence our results, we consider in this Paper, for simplicity, only 
the linear case $a=0$,  for which the
general solution to (\ref{e8}) can be written down  in the form
\begin{eqnarray}
&&\hspace{-4mm}\tilde \Sigma =  \exp ({-\beta  \tan^{-1} y/(2\tilde\Delta_s)})\left(C+{\tilde \Delta_s F\over 2}\right.\times\nonumber \\
&&\hspace{8mm} 
\left. \int_0^y dy^{'}
\exp ({-\beta  \tan^{-1} y'/(2\tilde\Delta_s)})\right),
 \label{e9}
\end{eqnarray}
where $y=2x/\tilde \Delta_s$ and $C$ is an integration constant.

In general the integral in (\ref{e9}) can be expressed in terms of
hypergeometric functions, but the resulting expression is
cumbersome  and difficult to use. The expression simplifies,
however, when $\tilde \Delta_s=0$. In this case we have
\begin{eqnarray}
 \hspace{-4mm}\tilde \Sigma &=& {\beta F\over 4}\left({1\over z}-e^z{\rm E}_1(z)\right ) + C_{+}e^{z} \quad {\rm  for}
\quad x > 0, \nonumber \\
\hspace{-4mm} \quad \tilde \Sigma  &=& {\beta F\over
4}\left (e^{-z}{\rm Ei}(z)-{1\over z}\right ) + C_{-} e^{-z}\quad {\rm for} \quad x < 0,
\label{e10}
\end{eqnarray}
where $z= \beta /( 4|x|), $
\begin{eqnarray}
{\rm E} _1(z)&=&\int^{\infty}_z {e^{-t}\over t} dt \quad
{\rm and}\nonumber\\
 {\rm Ei} (z)&=&\dashint^z_{-\infty} {e^{t}\over t} dt
 \end{eqnarray}
 When $ x
\rightarrow 0$, $z \rightarrow \infty$
  the term proportional
$C_{+}$ diverges exponentially. That means that we have to set
$C_{+}=0$. In this case we have $\tilde \Sigma (x=0)=0$.

When $\tilde \Delta_s\ne 0$ the expression for $\tilde \Sigma $ is
different from (\ref{e10}) with the most important qualitative
difference due to the fact that in this case  the  value of the
surface density at its minimum is non-zero. Assuming that the
position of the minimum,  $x = x_m < \tilde \Delta_s,$ from equation
(\ref{e8}) it follows that $\tilde \Sigma (x_m)\approx {\tilde
\Delta_s^2}F/\beta$. On the other hand $\tilde \Sigma (x_m)$
is much smaller than other terms in the expression
(\ref{e10}) when $x/\tilde \Delta_s \gg 1$, and, therefore, to
account approximately for a non-zero value of the minimum we simply
add $\tilde \Sigma (x_m)$ to the expression   given by  (\ref{e10}). In
this way we finally obtain
\begin{eqnarray}
 \tilde \Sigma &\approx& {\beta F\over 4}\left({1\over z}-e^z{\rm E}_1(z)+ 4{{\tilde \Delta_s}^2\over \beta^2}\right ) \quad {\rm for }
\quad x > 0, \quad \nonumber \\
\tilde \Sigma  &=&  {\beta F\over
4}\left (e^{-z}{\rm Ei}(z)-{1\over z} + 4{{\tilde \Delta_s}^2\over \beta^2}\right)+C_{-}e^{-z}
\hspace{1mm} {\rm  for}  \hspace{1mm}  x < 0. \label{e10aa}
\end{eqnarray}

The value of the
variable $z,$ corresponding to  the boundary values of $x$ 
of the zone  where the impulsive interactions operate,  
$x_{\pm}=\pm r_c /( 2r_{0}) ,$  namely 
 $z_{b}=\beta r_{0}/(2r_{c}),$  is expected to be large.
 Since $z \ge z_{b}$ throughout
the region $x_{-} \le x \le x_{+}$ when $z_{b} \gg 1,$ we can
simplify the expression (\ref{e10aa}) using the corresponding asymptotic
expressions of the functions ${\rm E}_1(z)$ and ${\rm Ei}(z)$ for large $z.$  In this way we
get
\begin{eqnarray}
\tilde \Sigma &\approx & {F\over \beta}(4x^2 + {\tilde \Delta_s}^2) \quad {\rm for} \quad x
> 0, \quad\nonumber\\ 
\tilde \Sigma &= & {F\over \beta}(4x^2 + 
{\tilde \Delta_s}^2)+C_{-}e^{-z} \quad {\rm for}  \quad x < 0,
\label{e12}
\end{eqnarray}
and we recall that $\tilde \Delta = (r-r_0)/r_0=2x$. 

From  (\ref{e12}) it follows  that when $F$ is fixed,
$\beta \rightarrow \infty$ and $\Delta_{s}\rightarrow 0, $ all terms in (\ref{e12}) tend to 
zero apart from the term, proportional to the constant $C_-,$  which   can be made
arbitrary large. This  means that  when  $q$ is large we expect the surface density at the
outer edge of the gap to be small compared to the surface density at the  inner edge
where there is a jump in $\tilde\Sigma$ of magnitude $C_-.$

However, as we shall see below,   the above analysis predicts a  minimum  surface density
in the gap that is too low. We investigate the possibility that this
is  because the scattering process is not entirely localised at one radial location
as has been assumed. This effect  is expected to have greater significance close
to the perturber.

\subsubsection{Modification of the gap profile close to the perturber}\label{minsig}

When the dimensionless distance from the perturber, $\tilde \Delta $, is sufficiently small, 
equation (\ref{e6}) may not be adequate for describing  the  surface density distribution.
We remark that  from equation (\ref{e3}) it follows that during one orbital period the quantity,  $\tilde \Delta,$ 
for a particular gas element changes according to
\begin{equation}
\delta \tilde \Delta ={2P_{orb}\dot J\over \sqrt{GMr_0}}=-{q^2\over \tilde \Delta^2
+\tilde \Delta_s^2}.
\label{g1}
\end{equation}
When $\tilde \Delta$ is sufficiently small, the magnitude of  $\delta \tilde \Delta$ can be of the order of 
the magnitude of $\tilde \Delta$ itself. In this situation a particular gas element crosses the perturber position 
 such that $\tilde \Delta< 0.$  Then
 every relative  period the surface density associated 
with this element changes by order of itself until
$|\tilde \Delta|$ becomes large enough so that
 $|\delta \tilde \Delta|$ becomes small compared to $|\tilde \Delta|.$

 Clearly, we  expect  a non-zero value of 
the surface density even when $\tilde \Delta_s=0$ and equation (\ref{e12}) predicts that $\tilde \Sigma
(\tilde \Delta=0)=0$. From the condition $|\tilde \Delta| < |\delta \tilde \Delta|$ we
have
\begin{equation}
\tilde \Delta < \tilde \Delta_*=q^{2/3},
\label{g2}
\end{equation} 
and the value of the  surface density at $\tilde \Delta=0$ may be simply estimated
with the help of equation (\ref{e12}) as 
\begin{equation}
\tilde \Sigma_{min}\equiv \tilde \Sigma
(\tilde \Delta=0)\sim \tilde \Sigma
(\tilde \Delta_*)\sim q^{4/3}{F\over \beta}.
\label{g3}
\end{equation} 
We comment that we have here simply assumed that the azimuthally averaged  $\Sigma$ does not decrease below the value
at  $\tilde \Delta = \tilde \Delta_*.$ This ignores azimuthally localised phenomena such as accretion onto the perturber
but this is not found to affect the azimuthally averaged profile  much in 2D simulations.

Let us consider this effect in more detail. We first introduce a new variable $y=q^{-2/3}\tilde \Delta$
and consider a map defined by

\begin{equation}
y_{n+1}=y_n -\frac{1}{ y_{n}^{2} +y_s^2},
\label{g4}
\end{equation}

\noindent where $y_s=q^{-2/3}\tilde \Delta_s.$   Here $y_n$ denotes the value of $y$ obtained after $n$ iterations,
the starting value being $y_0.$
 This map describes  successive changes to  $y$ for a gas
element occurring as a result of scattering,  obtained by application of  (\ref{g1}).
 The corresponding values  of the surface density  $\tilde  \Sigma_{n+1} $ and $\tilde  \Sigma_n$ are related through conservation 
of mass such that
 $\tilde \Sigma_{n+1} dy_{n+1}=\tilde \Sigma _n dy_n$:
\begin{equation}
\tilde \Sigma_{n+1}={\tilde \Sigma_n \over (1 +2y_n/(y_n^2+y_s^2)^2)}.
\label{g5}
\end{equation}

\begin{figure}
\resizebox{\hsize}{!}{\includegraphics{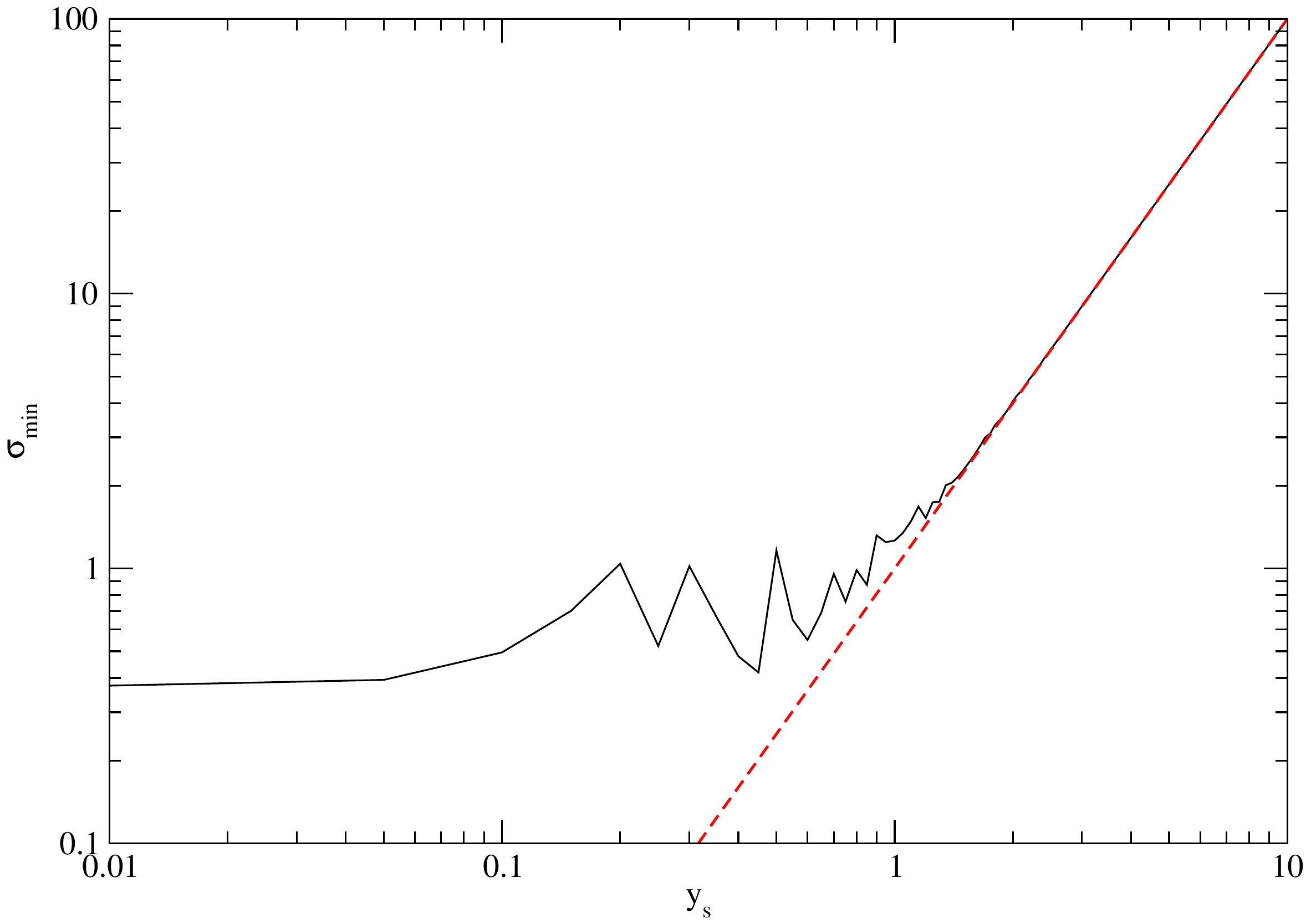}}
\caption{The dependence of the  mean ratio of the minimum surface density in the gap, to
the characteristic disk surface density $\Sigma_*$, $\sigma_{min},$ as a function of the parameter $y_s.$ 
The solid curve represents the result obtained from  the numerical iterations of the map
 based on the solution of equations (\ref{g4}) and (\ref{g5}),   while the dashed curve 
shows the analytic  expression $\sigma_{min}=y_s^2.$  \label{Fig0}}
\end{figure}

The map (\ref{g4}) and (\ref{g5}) can be iterated from some initial $y_{0}\gg 1$, where the distribution 
(\ref{e12}) is valid, and we have, accordingly,
\begin{equation}
\Sigma =\Sigma_* (y_{0}^2 + y_s^2), \quad \Sigma_*=q^{4/3}{F\Sigma_0\over \beta}.
\label{g6}
\end{equation} 
We iterate the map numerically starting from several initial values of $y_{0}$ to take into account 
a dependence of this procedure on initial conditions. They are chosen according to the rule:
$y_{0}=20+dy(i-1)$, $dy=0.01528\pi$ and $i=1,2,3..N$. An  iteration proceeds until  $y_n < 0$.
Then, we define a minimum  surface density corresponding to a particular $y_{0}$ as
$\Sigma^i_{min}=0.5(\Sigma_{n}+\Sigma_{n-1}).$  We then further average over the results obtained for  
different $y_{0}$. The result of this calculation is shown in Fig. \ref{Fig0}, where we plot the quantity
\begin{equation}
\sigma_{min} ={1\over N\Sigma_*}\sum_{i=1}^{N}\Sigma^i_{min}.
\label{g7}
\end{equation}
for $N=1000.$

As seen from  Fig. \ref{Fig0}  we can  approximate the dependence of $\sigma_{min}$ on $y_s$ as
\footnote{Clearly, there is some transitional zone in the region $0.1 < y_s < 2$, where $\sigma_{min}$ varies
in a complicated way. Since we need only a crude estimate of $\sigma_{min}$ we neglect this feature.}
\begin{equation}
\sigma_{min}(y_s < 1)\approx 0.4, \quad \sigma_{min}(y_s > 1)\approx y_s^2.
\label{g8}
\end{equation}
From equations (\ref{g6}) and (\ref{g8}), together with the definition of the parameter $\beta,$
we can obtain an estimate of the
minimum surface  density in the gap as
\begin{eqnarray}
\Sigma_{min}(y_s < 1)&\approx& \frac {1.2\pi q^{-2/3}F\nu_0\Sigma_0}{ \Omega_0 r_0^2}, \quad\nonumber \\
\Sigma_{min}(y_s > 1)&\approx& \frac{3\pi F\nu_0\Sigma_0}{\Omega_0 r_0^2}
 \left({{\tilde \Delta_s}\over q}\right)^2.
\label{g9}
\end{eqnarray}
We emphasise the approximate nature of these estimates which are found to be in general too small by a factor of a few
(see below). This is probably on account of the neglect of smoothing of the profile due to the action of pressure.  
Finally we would like to stress that although we effectively  assumed above that
the orbital radius of perturber $r_p$ is close to its initial value $r_0,$ the analytic  expressions are
valid for any $r_p,$  by  simply  replacing  $r_0$ by $r_p$ in them.
 \section{A simple procedure for calculating the  accretion disk 
evolution together with the  orbital evolution of the perturber occurring through torques exerted by the disk}\label{Simple}
As discussed above,  after the perturber
has been present in the disk  for a time that is larger
than its
characteristic evolution time scale,  but smaller
than the  characteristic time scale for  orbital evolution,
 the disk structure at radii,  $r\sim r_p,$ should be close
to a quasi-stationary one. 
In this situation,  the mass flux $\dot
M$ and 
the dimensionless value of  the specific angular momentum at the inner disk 
that appears in  equation
(\ref{en10a}) may be assumed to be functions of time only with a
characteristic time scale for change being much
larger than that required for  local disk  evolution.

On the other
hand, in the limit $q\ll 1$ the region in the vicinity of
perturber, where impulsive interaction operates, is very small,
with a typical  dimension  $\ll r_p$. Therefore, in the
simplest treatment of the problem we  describe the
influence of the perturber on the disk as providing  a jump condition on the surface density,  to be  applied
 at the perturber's orbital location,  in a disk otherwise evolving only under the influence of internal  viscosity.

As indicated above, the mass
flux through the gap is approximately constant in this limit. In
addition,  from  the steady state solution given by equation (\ref{en10a}) it can be seen that
when the mass flux is fixed, stationary solutions depend only on one constant, $h_*$, which is
proportional to the flux of angular momentum through the disk ${\dot L}={\dot M}\Omega_0r_0^2h_*. $

In Section \ref{s3} we showed   that the  outer disk
for which  $r > r_p$   should attain $\Sigma (r_{p+}) \sim 0$ (see discussion in the penultimate  paragraph of section \ref {gapstructure}).
This means
that the flux of angular momentum through the disk at radii $r >
r_p$ and $r\sim r_p$, $\dot L_{+}$, should be 
$\sim \dot M \sqrt{GMr_p}$ and we must set $h_{*}=\sqrt{r_p/r_0}$ in  equation (\ref{en10a}) in order  
for this to be applicable to  the outer
disk.

 On the other hand, the flux of angular momentum
through the inner disk, at radii $< r_p$, $\dot L_{-}$, should be
equal to the angular momentum accreted per unit time by the
component with the dominant mass, $M$.
 Assuming that $r_p$ is much larger
than the size of the last stable orbit of that component, we can
set $\dot L_{-}\approx 0$. Therefore, we set $h_*=0$ in  equation (\ref{en10a}) 
in order to apply that 
to the inner  disk located  at radii $r < r_p$. 

\noindent We  accordingly obtain 
\begin{eqnarray}
&&\hspace{-3mm}\Sigma(r)=
\left({\dot M(t)\over 3\pi \nu_*r_0^b}\right)^{1/( 1+a)} \hspace{-3mm}h^{-2b/( 1+a)},
 \quad {\rm for}\quad r < r_p 
\quad {\rm and}\nonumber
\\
&&\hspace{-3mm}\Sigma (r)={\left({\dot M(t)\over 3\pi \nu_*r_0^b}\right)}^{1/( 1+a)}\hspace{-3mm} 
h^{-(1+2b)/( 1+a)}\left(h-\sqrt{r_p/r_0}\right)^{1/( 1+a)}\nonumber \\
&& \hspace{-0.7cm} \quad {\rm for}\quad r > r_p.
\label{e13}
\end{eqnarray}
We remark that the first of these solutions appropriate to the outer disk  corresponds to a  steady state disk
with zero couple at $r = r_{p+},$ while the second solution appropriate to the inner disk corresponds
to a steady state disk with zero couple at a very small inner boundary radius.

\noindent For $r < r_p,$   equation  (\ref{e13}) gives   the
value of the surface density at the inner edge of the gap for the
linear case with $a=b=0$ as 
 \begin{equation}
 \Sigma (x=x_-)\approx {\dot M(t)\over 3\pi \nu_*}.
 \end{equation} 
This can be used to  obtain an  estimate the constant $C_-$ entering (\ref{e10}).
In particular,  in the limit $\beta \rightarrow \infty,$ for finite $F,$ 
$C_-$ becomes equal to  $\Sigma (x=x_-).$

We stress again that the solution (\ref{e13}) is approximately
valid only at scales such that  $r$ is of order $r_p.$  It clearly becomes invalid at a
length scale  which is large enough   that the characteristic time scale for evolution of  the disk
$t_{diff} $ becomes equal to or larger than the time $t$ after  which either the perturber  embedded in the disk,
or changes its orbital radius by an amount comparable to  $r_p.$
To calculate the disk evolution at large  radii it is necessary 
to use  equation (\ref{e6}) with the time dependence retained.

Since the total angular momentum of the
system is  conserved and there is no angular momentum flux through the
inner disk, the  outward angular momentum flux through the
outer disk, $T$, must be equal  and opposite to  the torque acting  
 on the perturber due to the disk, the latter thus  being $-T.$   We have, therefore,
\begin{equation}
T \approx - \dot M (t) \sqrt{GMr_p}
\label{e14}
\end{equation}
and we recall that as $ \dot M (t) > 0,$ $T < 0$ (see section \ref{s1.2}).

\subsection{Evolution of the surface density in the outer disk}\label{evoutd}
In order to model the evolution of the disk surface density together with the
orbital evolution of the binary,  we implement a procedure that   updates the disk surface density
using 
equation (\ref{e6}) with the torque terms corresponding to interaction with the disk
being set to zero. 

 In our analytical work we have so far assumed, for simplicity, 
that the binary semi-major axis, $r_p$, is close to its initial position,
$r_{0}$. However, many expressions,  such as e.g. the distribution of the surface density in the gap,  remain approximately 
valid even when $r_p$ is noticeably smaller than $r_0,$ provided that we substitute $r_p$ for  $r_0.$  This
can be understood as follows. When the perturber 
is sufficiently heavy,  from the discussion in the previous section, the solution close to its orbit
is  quasi-stationary,  such that the fact that at any time   $r_p$  is changing,  plays only a minor role.  In this situation,
it is convenient to change the unit of length in (\ref{e6}) from $r_0$ to $r_p,$ neglecting  $\dot r_p$ 
when carrying out this transformation.  Accordingly, 
from now on we switch  the unit of length from $r_0$ to $r_p$ in the definition of
the variable $h$ in (\ref{e6}) so that   $h=\sqrt{r/r_p}.$ This  coincides
with the previous definition $h=\sqrt{r/r_0}$ only when $t=0.$ Other quantities are appropriately rescaled
apart from 
those involving 
$\Sigma_0, $ which as before   denotes  the initial surface density at $r=r_0.$
Then we can  continue to  use the kinematic viscosity prescription defined through  (\ref{e5}).
We further remark that when these  changes  are  made, 
the second term in brackets in the second expression in (\ref{e13}) is equal to unity.

Under the above  conditions equation (\ref{e6}) gives the equation for the evolution of
the surface density as
\begin{equation}
{\partial  \tilde \Sigma  \over \partial \tau}=  {1\over h^3}
{\partial^2 \Phi\over \partial h^2}\quad {\rm with}\quad 
\Phi=h^{1+2b} {\tilde \Sigma}^{1+a}. 
\label{e15}
\end{equation}
in the region $h > 1$ with the initial conditions defined by equation
(\ref{en9}) and the inner boundary condition\\  $\Tilde \Sigma(h=1)=\Phi (h=1)=0.$
\footnote{Note that in fact the initial distribution (\ref{en9}) is not compatible 
with the inner boundary condition, accordingly, in practice, when solving (\ref{e15}) numerically
we modify the stationary solution (\ref{en9}) by adjusting the  surface density profile 
within a small transitional region $1 \le h \le 1+\Delta h$, $\Delta h < 1,$  in order  that $\tilde \Sigma (h=1)=0.$ 
An exact form of
the surface density profile in the transitional region is not important for the solution at large time 
$\tau > 1$.}. 
\noindent    
We adopt an  outer boundary condition that  either corresponds to a disk of formally infinite extent
or corresponds to  a finite boundary at an outer radial distance, taken to be $h_{out}=\sqrt{5}$. In the former case 
we assume that asymptotically, when $h \rightarrow \infty$ 
the angular momentum flux tends to zero, i.e. the disk approaches the solution (\ref{en9}) with $F_0=1$. Then,
we find the mass flux using equation (\ref{en8}) with $F=\partial \Phi / \partial h .$    In the latter case
we assume that there is no mass flux through the outer boundary at $h_{out}=\sqrt{5}$. The results obtained with 
this boundary condition can be directly  compared with  numerical simulations, most of which adopted this boundary condition.

\noindent  While  the surface density is being updated  using the above procedure, we use equations (\ref{en1}), (\ref{en2}) and 
(\ref{e14}) to simultaneously  find the orbital evolution of the perturber. 
However, note that after the unit of length has been changed from $r_0$ to $r_p,$ the value of  $h_{out}$ corresponding to a fixed 
pre-scaled radius grows with time. The effect of this can be taken into account  within the framework of our approximation scheme by use
of a grid of solutions to (\ref{e15}) corresponding to different $h_{out}.$   The situation can be further simplified for
the case of a constant kinematic viscosity when equation (\ref{e15}) becomes linear. In this case, its late time solution
is mainly determined by the smallest eigenvalue of the  linear eigenvalue problem determining the associated normal modes. 
The  dependence  of this on $h_{out}$ is discussed in the Appendix 
where it is illustrated in Fig. \ref{Fig2}. In our numerical work we have only considered modest changes of $r_p$  for which this effect
turns out to be unimportant.  It is not, accordingly, taken into account in the rest of the Paper.   

\begin{figure}
\resizebox{\hsize}{!}{\includegraphics{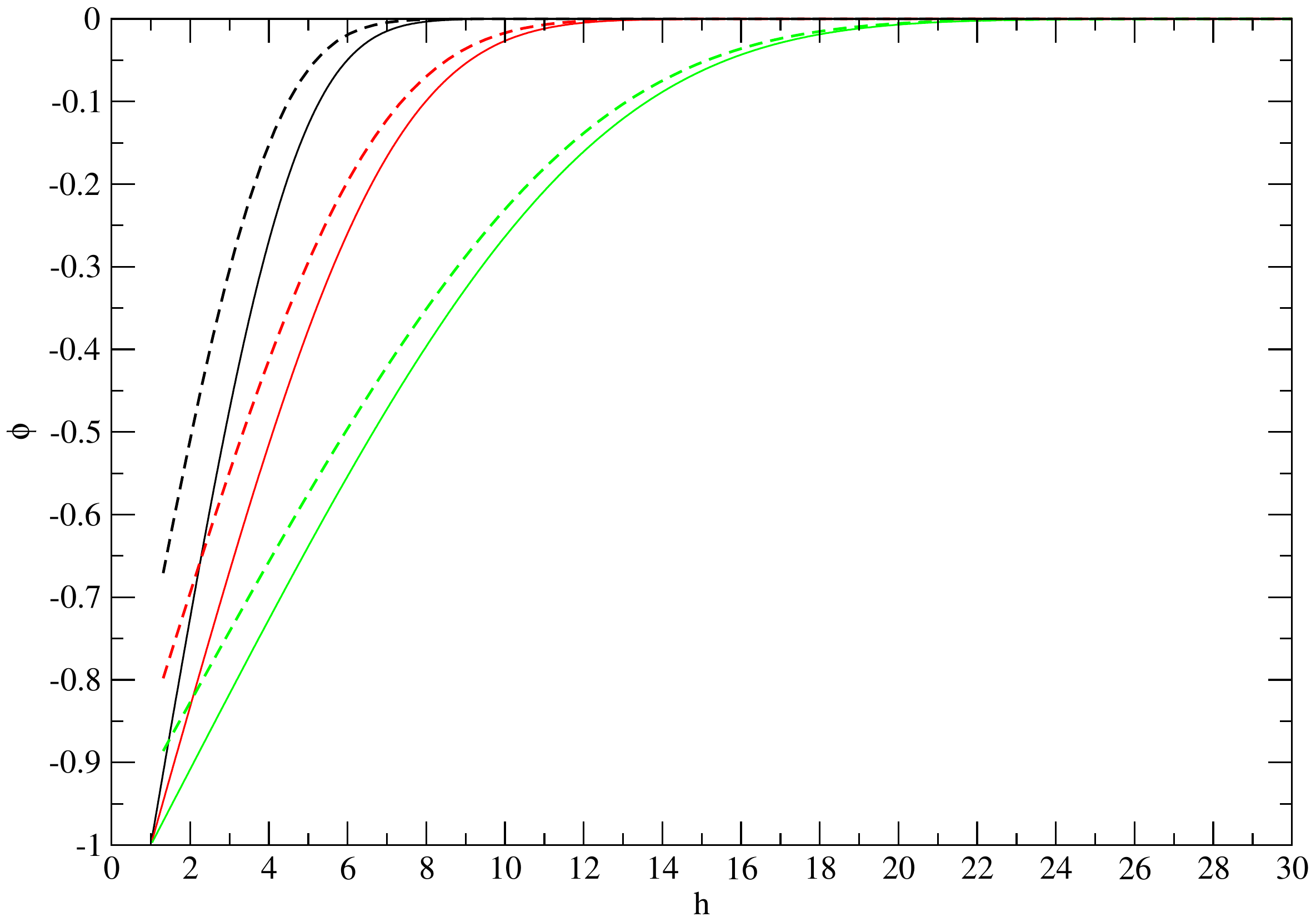}}
\caption{The difference $\phi = \Phi -h$ is shown as a function of the radial coordinate $h$,
at  $\tau=5$, $20$ and $100$,  see the text for details.
} \label{phi}
\end{figure}

\begin{figure}
\resizebox{\hsize}{!}{\includegraphics{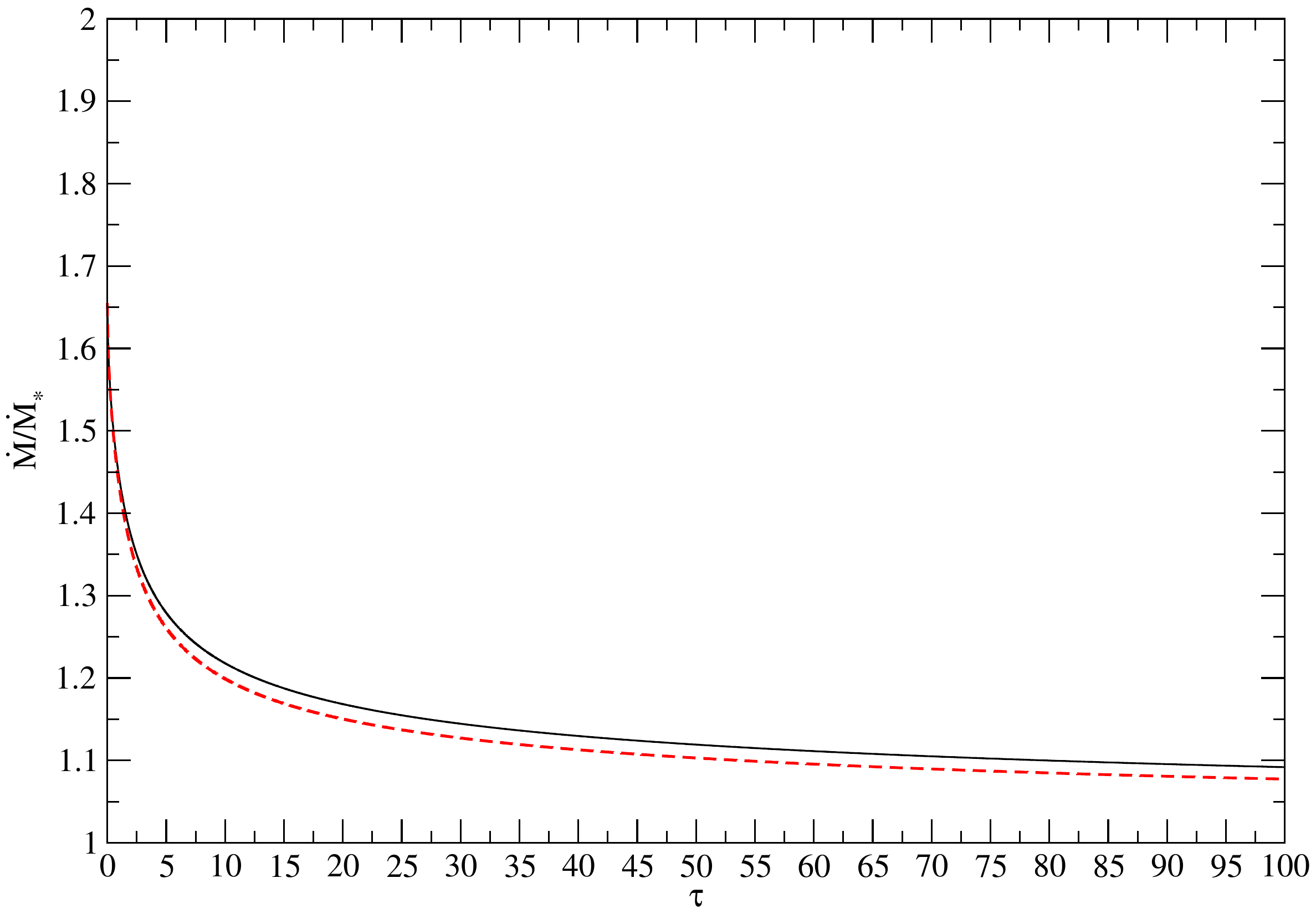}}
\caption{The dimensionless mass flux $\dot M/\dot M_*$ is shown as a function of time $\tau$. 
The solid and dashed curves correspond to $a=2/3$, $b=1$ and $a=3/7$, $b=15/14$, respectively. 
} \label{mdot}
\end{figure}

\subsection{A similarity solution for an  accretion disk of  infinite extent}
\label{infdisc}

For arbitrary $a$ and $b$, 
equation (\ref{e15}) has in general  to be solved numerically. However, simple arguments allow us to show that when
 $\tau \gg 1$ the accretion rate close to the perturber is approximately equal to that at infinity.
In order to find the corresponding approximate non-stationary 
solution to (\ref{e15})  we note that it must have the 
property that 
\begin{eqnarray}
\Phi(r)& \sim & h-1, \quad {\rm for} \hspace{2mm}   r < r_{diff}(\tau) \quad {\rm and}\nonumber\\
 \Phi (r) & \rightarrow h& \quad  {\rm for} \quad r \rightarrow \infty 
\label{e16}
\end{eqnarray}
to satisfy the inner boundary condition and the requirement that the disk tends asymptotically to
the stationary solution with zero angular momentum flux.  We recall that $ r_{diff}(\tau)$ is the radius at which the local
viscous diffusion time is equal to the current time $\tau$ in dimensionless form.  Assuming that $r_{diff} \gg r_p$ the
difference $\phi =\Phi -h$ should be small at $r > r_{diff}$ and we can treat it as a perturbation. In this
case we  can obtain a linear equation for $\phi$  by linearizing  (\ref{e15}). This takes the form       
\begin{equation}
{\partial  \phi \over \partial \tau} =  (a+1) h^{{2(b-(a+1))\over a+1}}
{\partial^2  \phi \over \partial h^2}. 
\label{e17}
\end{equation}

\begin{figure*}
\centering
\includegraphics[width=17cm]{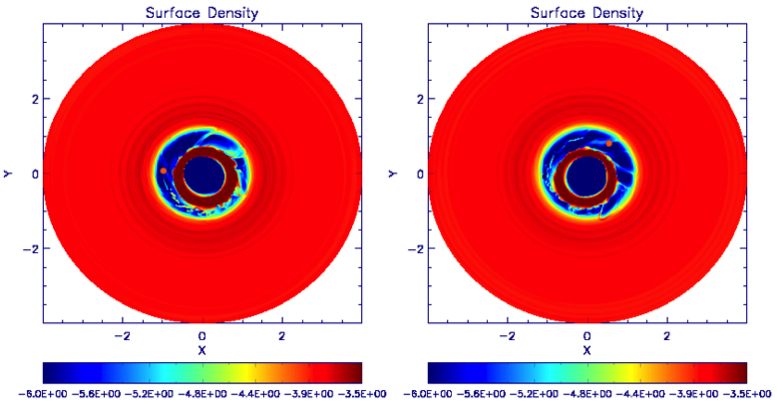}
\caption{  $\log \Sigma$  contours for $q=0.02$ with softening length $0.1H$
 after $50$ orbits (left panel) and after $100$ orbits (right panel).
In these simulations the companion, its position in each case being at the centre of  the small  red circle located within the gap region,  was allowed to accrete. 
 The width of the gaps  slowly increase  while the accretion rates,  on average,  slowly decrease  with time.
Short wavelength density waves  in the outer disks are  just visible. Note that values of  $\log \Sigma$ below the minimum indicated on the colour bar are plotted as that minimum value  
} \label{2e-2100P}
\end{figure*}

\begin{figure*}
\centering
\includegraphics[width=17cm]{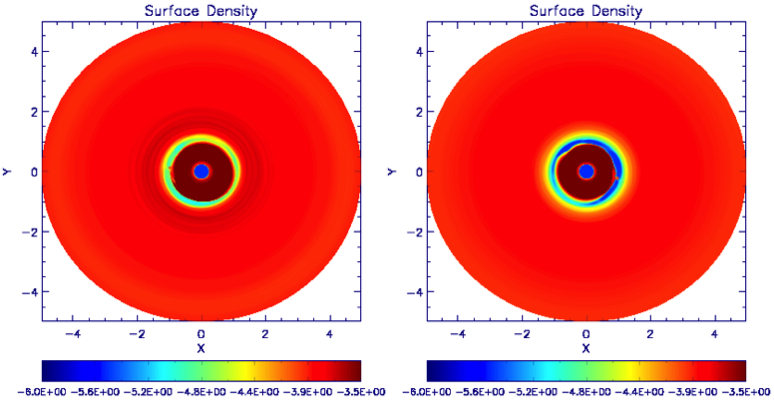}
\caption{   As in Fig. \ref{2e-2100P} but for $q=0.01$ with softening length $0.6H$  after $100$ orbits (left panel )
 and $800$ orbits (right panel).
As the mass ratio is lower  in this case compared to that  of Fig. \ref{2e-2100P} the gap in the disk is narrower. The companion, indicated by
a small red circle  is found in general to  orbit closer  to the inner disk edge
at earlier times. In the left hand panel the companion grazes the inner edge slightly above  the $x$ axis for 
$x < 0.$
This  enhances the accretion rate at that stage.
} \label{1e-2100P}
\end{figure*}

We look for a self-similar solution to (\ref{e17}) of the form $\phi=\phi(\xi)$, 
where the similarity variable $\xi = h\tau^{-\gamma}$ and  
 $\gamma~=~(a+1)/2c$,  with  $c=2(a+1)-b.$ The function  $\phi$ then satisfies
 the ordinary differential equation  
\begin{equation}
{d^2 \phi \over d\xi^2} +{1\over 2c}\xi^{{3(a+1)-2b\over a+1}}{d \phi \over d\xi}=0
\label{e18}
\end{equation} 
and we assume from now on that $c > 0$.   Note that $r_{diff}(\tau)$ can be found from the requirement 
$\xi(r_{diff},\tau)=1$ and we, accordingly, have $r_{diff}=r_p\tau^{2\gamma}$.
The general solution to (\ref{e18}) is
\begin{equation}
\phi=C_1 + C_2\int_{0}^{\xi}d\eta \exp{(-(a+1)\eta^{{2c\over a+1}}/(4c^2))}. 
\label{e19}
\end{equation}
where $C_1$ and $C_2$ are constants of integration.  These are determined by the requirement that
the solution should satisfy the conditions that $\phi \rightarrow -1$ for $\xi \rightarrow 0$ and 
$\phi  \rightarrow 0$ for $\xi \rightarrow \infty$\footnote{Note that the inner boundary condition formally assumes 
that $r_p\approx 0$ and is approximately valid only when $r_{diff} \gg r_p$.}. These ensure that the surface density
profile matches the   required forms for $h=1$ as $\tau \rightarrow \infty,$  and  $h \rightarrow \infty$ for all $\tau.$ 
  We therefore find
\begin{equation}
\phi={1\over I}\int_{0}^{\xi}d\eta \exp{(-(a+1)\eta^{{2c\over a+1}}/(4c^2))}-1, 
\label{e20}
\end{equation}  
where
\begin{eqnarray} 
I&=&\int_{0}^{\infty}d\eta \exp{(-(a+1)\eta^{{2c\over a+1}}/(4c^2))}\nonumber \\
&=&{(a+1)\over 2c}\left({a+1\over 4c^2}\right )^{-{(a+1)\over2 c}}
\Gamma ({(a+1)/2 c})
\end{eqnarray}
with  $\Gamma$ denoting  the gamma function. 
 
In order to demonstrate the applicability of  the similarity solution described above,
we illustrate  a  numerical solution  of equation (\ref{e15}) in Figs. \ref{phi} and \ref{mdot}.
In Fig. \ref{phi} the quantity $\phi = \Phi - h$ is shown at 
times  $\tau=5$, $20$, $100$ with curves passing through  larger values  at the same $h$ corresponding to 
later times. The solid curves are obtained by numerical solution of equation (\ref{e15}) while the dashed curves 
represent the analytical expression (\ref{e20}).  
For the solution shown,
$a=2/3$ and $b=1$ which are the values expected for a disk with Thompson opacity being dominant.
For  initial condition we took  $\Phi=h$,  with the  adjustment as described in section \ref{evoutd} , which corresponds 
to the distribution given by  (\ref{en9}) with $F_0=1.$
 
One can see  that the quantity $\phi$  does have
the form predicted by the expression (\ref{e20}).
 It tends to $-1$ at small values of $h$ and to zero at large values
of $h$ with intermediate values being gradually shifted towards larger $h$ with time. 
In Fig. \ref{mdot} the dimensionless
mass flux $\dot M/\dot M_*,$  where
$\dot M_*  =3\pi \Sigma_0 \nu_0 ,$ is shown as a function of time for the case  $a=2/3$, $b=1$ (solid curve) and the case
of $a=3/7$ and $b=15/14$, the latter  corresponding to a disk  opacity dominated by free-free transitions.
 It is clear that the mass
flux tends asymptotically to its unperturbed value when $\tau \rightarrow \infty$. It can be also seen that the ratio 
$\dot M/\dot M_*$ never exceeds $\sim 1.6$ for all times.

\section{Orbital evolution time scale}\label{orbtim}

Since the time dependent mass flux never significantly exceeds its unperturbed value, and tends asymptotically to this value,
we estimate the mass flux in (\ref{e14}) as $\dot M(t)\approx \dot M$, where $\dot M$ is the unperturbed mass flux, i.e. the one that 
existed in the disk before the secondary had been embedded. Substituting the result for  $T$ in equations (\ref{en1}) and
(\ref{en2}) we get  
\begin{equation}
r_p=r_0\exp{(-t/t_{ev})} , \quad t_{ev}={M_p\over 2\dot M}.
\label{e21}
\end{equation}  
Thus, in this case the orbital evolution time scale is equal to the time that would be required for  the steady state   accretion flow through the unperturbed disk
to amount to   half of the perturber mass, $M_p/2$.  
We comment that  this evolution time scale was  given by   Nixon et al. (2011) and  Roedig et al. (2014) ( see  equation (9) in  the latter paper
in  the limit of small mass ratio and small eccentricity with no accretion onto the perturber ).
This was obtained by simply assuming that  mass flowing through the orbit transfers  {\it all} its angular momentum to the companion.   Here  we have considered   
the process of   gap formation,  which is necessary for this time scale to be valid,   as well  as demonstrated  
compatibility with the evolution of the accretion disk which supplies the mass flow 
via material on approximately circular orbits. 

\section{Estimating  the accretion rate onto the secondary}\label{est}
We now  give simple estimates for the possible accretion rate  onto the perturber in the spirit of  our simplified 1D modelling approach.
  concluding that this is not expected to produce significant
effects on orbital evolution.  However, it can lead to  non-trivial activity and  be relevant to studies aimed at the
investigation of  means of detection of SBBH and other similar systems.

In order to estimate the accretion rate, $\dot m$, to the secondary we assume that all gas elements, which approach the secondary within an accretion radius, $r_a$, are
accreted to the secondary. $r_a$ is estimated as in  standard Bondi-Hoyle accretion 
(Bondi \& Hoyle 1944),  but we take into account the fact that  a typical relative velocity of a gas element with respect 
to the secondary, $v_{rel}$, 
is twice of the Keplerian value. Accordingly we obtain  $r_a={q}r_p/2.$
Adopting these assumptions the accretion rate to the secondary, $\dot m$, can be easily estimated 
in two limiting cases $r_{a} \gg H$ and $r_{a} \ll H$
\footnote{Note that the latter case formally violates the sufficient condition 
for a  'sufficiently massive' binary (\ref{en13}). Nonetheless we consider it to compare our analytical 
estimates with the results of numerical simulations, for which the disk is relatively thick, and which  marginally corresponds to this case even though
$q$ exceeds both $q_1$ and $q_2$  so that the binary is  in fact 'sufficiently massive'.} as  
\begin{eqnarray}
\dot m_1\approx 4r_aH\rho_{min}v_{rel} \quad \dot m_2\approx \pi r_a^2\rho_{min}v_{rel},
\label{e22}
\end{eqnarray}
where the indices $1$ and $2$ correspond to the former and latter cases, respectively, $\rho_{min}\approx \Sigma_{min}/(2H)$ and 
we use the first expression in (\ref{g9}) to estimate $\Sigma_{min},$  making the  assumption  that the softening length is sufficiently 
small. Using the explicit expressions of all quantities 
entering (\ref{e22}), we obtain
\begin{equation}
\dot m_1\approx 4\pi q \left ({\Sigma_{min}r_0^2\over P_{0}}\right) \hspace{1mm}{\rm and}\hspace{1mm} \dot m_2\approx {\pi^2\over 2}{q^2\over \delta}\left ({\Sigma_{min}r_0^2\over P_{0}}\right),
\label{e23}
\end{equation}
where we have assumed   for simplicity, that $r_p \sim r_0$.

We now use  the estimate of the minimum surface density given by  (\ref{g9}) in (\ref{e23})  thus  obtaining
\begin{equation}
\dot m_1\approx 2.5\pi q^{1/3}F\nu_0\Sigma_0, \hspace{1mm} {\rm and}\hspace{1mm} \dot m_2\approx 0.3\pi^2 q^{4/3}\delta^{-1}F\nu_0\Sigma_0
\label{e24}
\end{equation}
The expressions (\ref{e24}) can be further simplified for a disk of infinite extent. From the discussion
in Section \ref{infdisc} it follows that in this case  $F\sim 1$ and, accordingly, the accretion rate through 
the disk $\dot M \sim \dot M_*=3\pi \Sigma_0 \nu_0.$  Therefore, from the first expression in (\ref{e24}) we expect 
that 
\begin{equation}
\dot m\approx q^{1/3}\dot M,
\label{e24n}
\end{equation}
where we have neglect a numerical factor order of unity.
It then follows from the expression (\ref{e21}) for the orbital evolution time that the mass of the secondary
can only increase by  a fraction $q^{1/3}/2$ during significant orbital evolution, even when the disk is very thin.  
This fraction will be  smaller for a thick disk.  We remark that 
in addition, the disk could be thickened locally as a result of radiative processes,  producing a back reaction on  the accretion flow.
 Thus  accretion is not expected to affect the orbital evolution significantly. 
It is important to stress that although we have assumed that $r_{p}\sim r_0$ when deriving (\ref{e24n}), this expression
is, in fact, valid for any $r_{p}$. 

\section{Numerical simulations of gap formation, migration and accretion onto the perturber}\label{Sims}
In this section we  consider  numerical simulations that resulted in gap formation. They  were performed  for 
$q=0.01$ with  standard softening,
and for $q=0.02$ and $q=0.01$ with  small softening (see Section \ref {Esim}).
For cases with standard softening, the computational domain
was taken to be $(0.2r_0, 5r_0).$  For the other  cases it  was $(0.4r_0, 4r_0).$ 
Other features of the simulations were as described in Section \ref{Esim}.

\subsection{Accretion and migration}\label{Acmig}
Some simulations were performed with the perturber allowed to accrete mass from the disk.
Following the approach of Bondi \& Hoyle (1944),
the accretion rate is then taken to be ${\dot  m} = \pi r_{a}^2\Sigma v_{rel}/(2H)$ (see equation (\ref{e22}) ).
In implementing this we removed mass  uniformly from the grid cell containing the perturber and its eight nearest neighbours.
In this context we remark that in the small softening cases the grid size is approximately equal to, or somewhat less than,
both the softening length and the accretion radius. Furthermore tests showed that results in particular for the total amount of mass accreted
over long periods of time, were not sensitive to the detailed implementation,
not being significantly changed even  when the mass was removed  only from  the grid cell containing the perturber.

To calculate the torque acting on the perturber, contributions arising from gravitational torques
from the disk as well as the mass and  momentum accreted by the perturber were taken into account.
The latter effects were found to play only a minor role for the simulations reported here.
As indicated above simulations with $q  <  \sim 0.003$  did not readily exhibit gap formation,
with the perturber remaining in a type I  migration regime,  at least  on time scales of a few hundred orbits,
although this picture could potentially be modified  on very long time scales through the cumulative dissipation of weak shock waves.

The structure of the disk  gaps  for $q=0.02$ and $q=0.01$  is illustrated in the surface density contour plots presented  in
Figs.  \ref{2e-2100P} and  \ref{1e-2100P} at various times.  The runs illustrated respectively
correspond to the strongest and weakest gap forming
cases considered in this section.  Note that the gap is indeed significantly wider and deeper for $q=0.02$ as expected and in addition  the gap edges
 define  significantly non circular boundaries as mentioned above. Material crossing the gap in the form of streamers is also present. Note that animation of the process of gap formation can be found on website
http://astro.qmul.ac.uk/people/sijme-jan-paardekooper/publications.

\begin{figure}
\resizebox{\hsize}{!}{\includegraphics[angle=270]{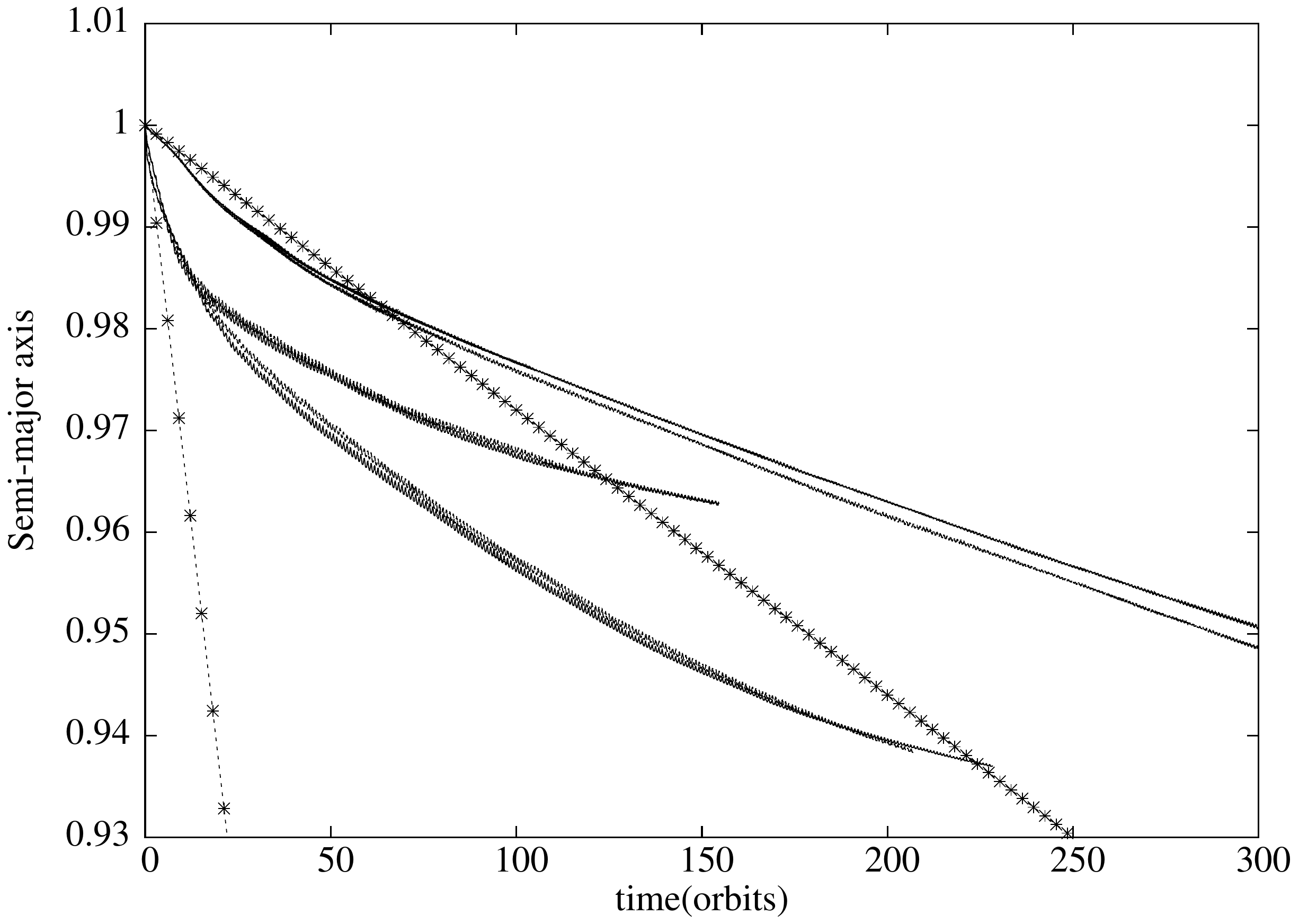}}
\caption{ Semi-major axis, in units of the initial orbital radius, as a function of time for $q=0.02$ and $q=0.01$ for small softening
and  for $q=0.01$ with standard softening. Two curves without imposed crosses,  which are very close together,  are shown for each of these three cases.
The uppermost pair of curves corresponds to $q=0.01$ with standard softening and the lowermost pair
for  $q=0.01$ with small  softening. The central pair corresponds to  $q=0.02$ with small softening.
The lower of the  pair of curves  for the cases with small softening correspond to  runs  with accretion  from the disk included (see text).
For the case with standard softening this situation is reversed.
The straight lines with imposed crosses is obtained by  adopting the initial type I  migration rate
with contributions from all  $m\le 100$ included. The line with the more widely separated crosses corresponds to $q=0.01$ with
small softening while the other line corresponds to $q=0.01$ with standard softening.}
 \label{Migration.eps}
\end{figure}

The semi-major axis is shown  as a function of time for $q=0.02$ and $q=0.01$ for small softening
and  for $q=0.01$ with standard softening in Fig. \ref{Migration.eps}.
The behaviour depends only very weakly on whether the perturber is allowed to accrete from the disk or  not.
  At early times the cases   with $q=0.01$   have the   migration rates expected in the type I regime  (see Fig. \ref{Migration.eps})
 being approximately
  $\propto \Delta_s^{-1}$ in agreement
with the discussion  in section (\ref{typeirates}) above. However, after a few orbits the effects of gap formation become noticeable
 and the migration starts to  slow down.
For the case with $q=0.02$  the initial migration rate is a factor of two smaller than the expected type I
migration rate with the effects of gap formation being  noticeable immediately.
Note that at longer times the migration rates for  $q=0.01$ with different softening lengths
 slow to become approximately equal as would be expected
if the migration was governed by the viscous evolution of the disk. There was  a deeper and wider gap in the small softening case,
the ultimate  migration  being  somewhat slower. This might be expected on account of the wave excitation and
associated  angular momentum removal being  apparently more effective
in  the case with standard softening.

\begin{figure}
\resizebox{\hsize}{!}{\includegraphics[angle=270]{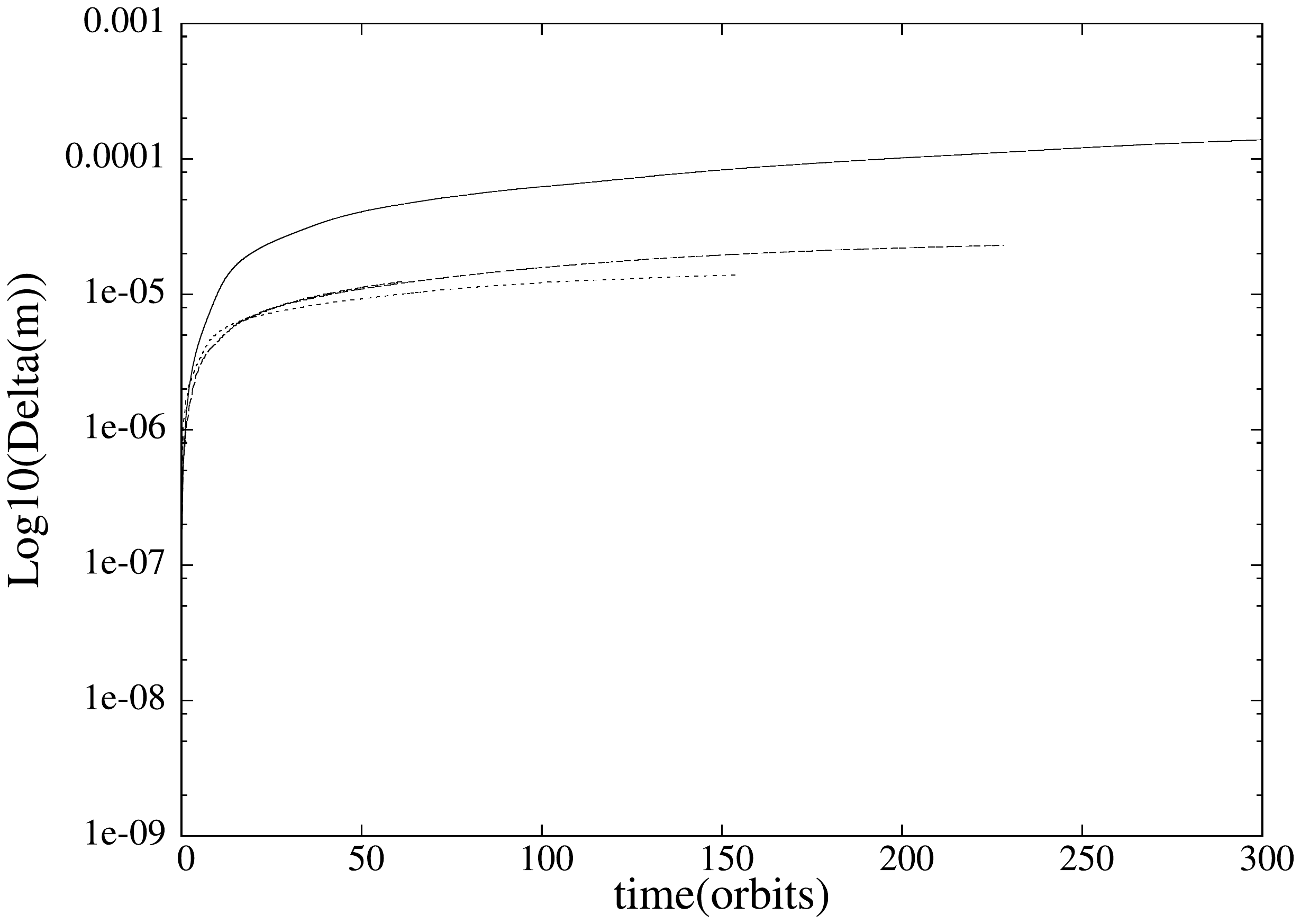}}
\caption{  The mass accreted, in units of $M$,  as a function of time for  runs with accretion illustrated in Fig. \ref{Migration.eps}.
The uppermost curve corresponds to  $q=0.01$ with standard softening,  the central curve
to  $q=0.01$ with small  softening and the lowermost curve corresponds to   $q=0.02$ with small softening. Note that although
the accretion rate shows large variations, these curves  do not because
they effectively time average this quantity. }
\label{Accretion.eps}
\end{figure}
 
On the other hand,  the larger open  inner boundary radius adopted for
the  simulations with smaller softening, on account of necessary  numerical convenience, results in
a relatively larger angular momentum loss as material passes through it.

 This not available to the secondary mass
resulting in an expected  slowing of the the inward migration rate
as compared to simulations with a smaller inner boundary radius (see discussion in section \ref{Simple} below).
 The case with $q=0.02$ ends with slowest migration rate as expected.
 In all cases the characteristic time scale becomes  comparable  to or greater than that for the viscous evolution of the disk.
To consider this issue more precisely a comparison between the orbital evolution obtained from  some of these simulations
and the semi-analytic model for which the orbital evolution is driven by the viscous evolution of the disk
is given in section \ref{orbtim}.

Accretion onto the secondary from the disk plays only a minor role in these simulations as illustrated in Fig. \ref{Accretion.eps}.
We remark that the accretion rate can show large variations on an orbital time scale. These are due to the disk edges not being circular
enabling the secondary to approach,  in particular the inner disk edge,   where the accretion rate naturally increases, quasi-periodically on an orbital time scale. 
In spite of this it is found that the time averaged  accretion rate is well behaved and significantly slower for cases with wider and deeper gaps as expected.
This is discussed further in section  \ref{compsim} below.
For the simulations presented here the accretion rate is never large enough to double the perturber's mass
on its migration time scale.

\subsection{Comparison of analytical and numerical results on the gap profile}\label{comp gap}

\begin{figure}
\resizebox{\hsize}{!}{\includegraphics{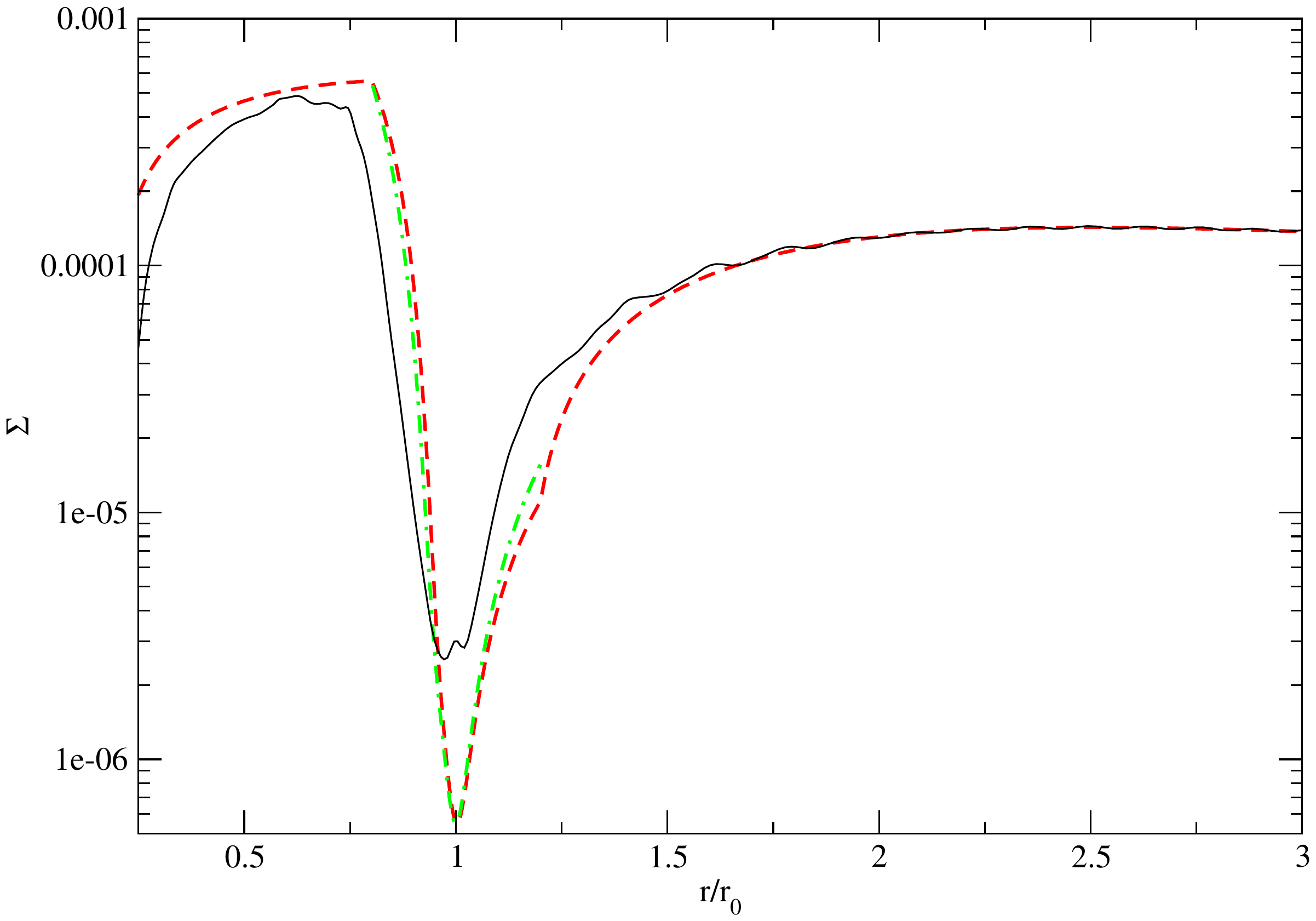}}
\caption{The surface density $\Sigma $ in units
$M/r_0^2$ is shown as a function of the radial coordinate, for the
moment of time corresponding to $t=1500P_0.$
The perturber was initiated in the disk at $t=0.$
The solid curve is obtained from 
a direct two dimensional  simulation and averaging over the
azimuthal coordinate. The dashed curve is obtained from  a numerical solution of
equation (\ref{e6}) and the dot dashed curve gives the shape of
the gap according to equation (\ref{e10}) with the constants $F$
and $C_{-}$ obtained from  a fit to  the numerical solution of (\ref{e6}), 
see the text for more details.} \label{Fig1}
\end{figure}
We compare results of a  numerical   simulation of the type 
described in Section \ref{Sims}
with results obtained from the  numerical solution of
equation (\ref{e6}) within the framework of the  simplified one-dimensional model and also
with an analytic surface density profile obtained (\ref{e10aa})  in Fig. \ref{Fig1}.
The simulation adopted corresponds to the case with the mass
ratio  $q=10^{-2}$ and standard softening described in Section \ref{Sims}.
 The initial surface density profile   was
taken to be $\Sigma = \Sigma_0(r/r_0)^{-1/2}$ which
  corresponds to
$F_0=0$ in equations  (\ref{en9}) and  (\ref{en10}). 
The value  $\Sigma_0=2.4\times 10^{-4}M/r_0^2$ was adopted to correspond to the
numerical simulation. 

\noindent In Fig.  \ref{Fig1} the surface density $\Sigma $ is shown as a
function of $r/r_0$ for  $t=1500 P_0$,  we recall 
that $P_0=2\pi \sqrt {r_0^3/ GM}.$  Note that we  determine the constants $F$ and
$C_-$ in (\ref{e10aa}) using the solution  obtained with help of our one-dimensional numerical approach,
by matching the values of mass flux through the gap and a value of the surface 
density at the inner edge of the gap.
We see from Fig. \ref{Fig1} that  the profiles of
$\Sigma $ given by the analytical expression (\ref{e10aa}) and the
numerical solution of equation (\ref{e6}) almost coincide. 
There is also a good agreement between  the surface density profiles obtained 
from these approaches and those obtained from  two
dimensional numerical simulations apart from the region very close to the
surface density minimum, In this region  the two dimensional simulations  give
values of $\Sigma $ roughly five times larger than the  other approaches.
This can be explained by having adopted an oversimplified treatment of the angular momentum exchange 
within the framework of the impulsive approximation.  Also note that the estimate
of the minimum surface density based on equation (\ref{g9}) gives practically the same 
values as that obtained from (\ref{e10aa}), and, therefore, is not shown.

\begin{figure}
\resizebox{\hsize}{!}{\includegraphics{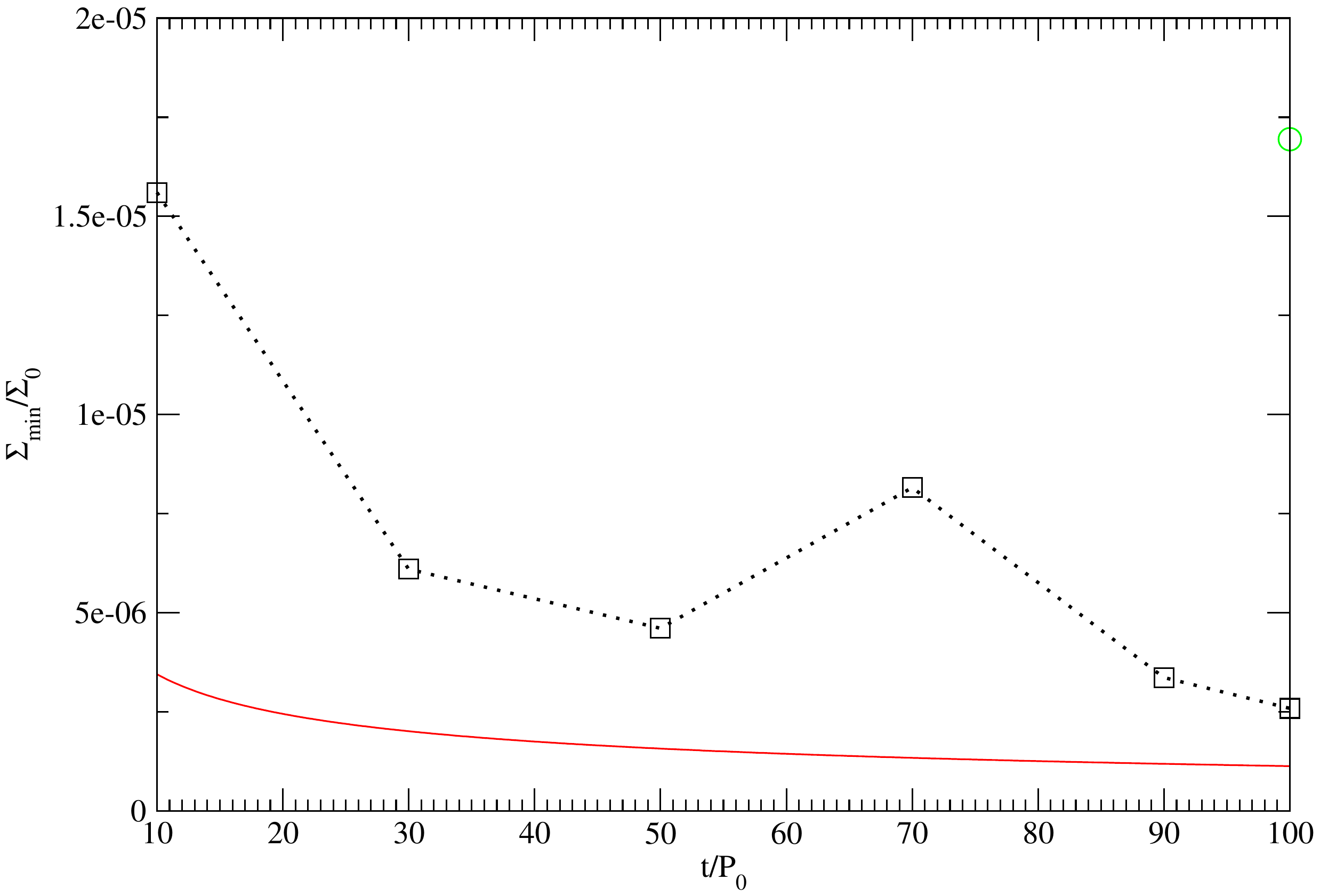}}
\caption{The minimal value of surface density $\Sigma_{min}$ in the gap in units
of initial surface density $\Sigma_{0}=2\cdot 10^{-4}M/r_0^2$ is shown as a function of the dimensionless time 
$t/P_{0}$. Squares connected by the dotted line represent the result of two-dimensional
numerical calculations while the solid curve is calculated according to the first 
expression (\ref{g9}).} \label{Fig1aa}
\end{figure}

In that context in  Fig. \ref{Fig1aa}  we compare  numerical simulation  results for the minimum surface 
density in the gap with the expression (\ref{g9}). The parameters of  the numerical simulation we  consider 
are the same as described above except that the mass ratio is $q=2\cdot 10^{-2}$
and $\tilde \Delta_s=5\cdot 10^{-3}.$  The value of the dimensionless flux 
$F$ entering (\ref{g9}) is calculated  as outlined  in Appendix \ref{disc_ev}.
As seen from  Fig. \ref{Fig1aa}, apart from the at the  times $t=  10P_0$ and 
$t = 70P_0$, where the analytical results give much smaller value of $\Sigma_{min},$ 
the analytical and numerical results are in reasonable agreement with the numerical results
giving $2-3$ times larger values of $\Sigma_{min}$  as indicated in section \ref{minsig}.

\subsection{Comparison of the analytic estimate of the accretion rate  with the results of a simulation}\label{compsim}

\begin{figure}
\resizebox{\hsize}{!}{\includegraphics{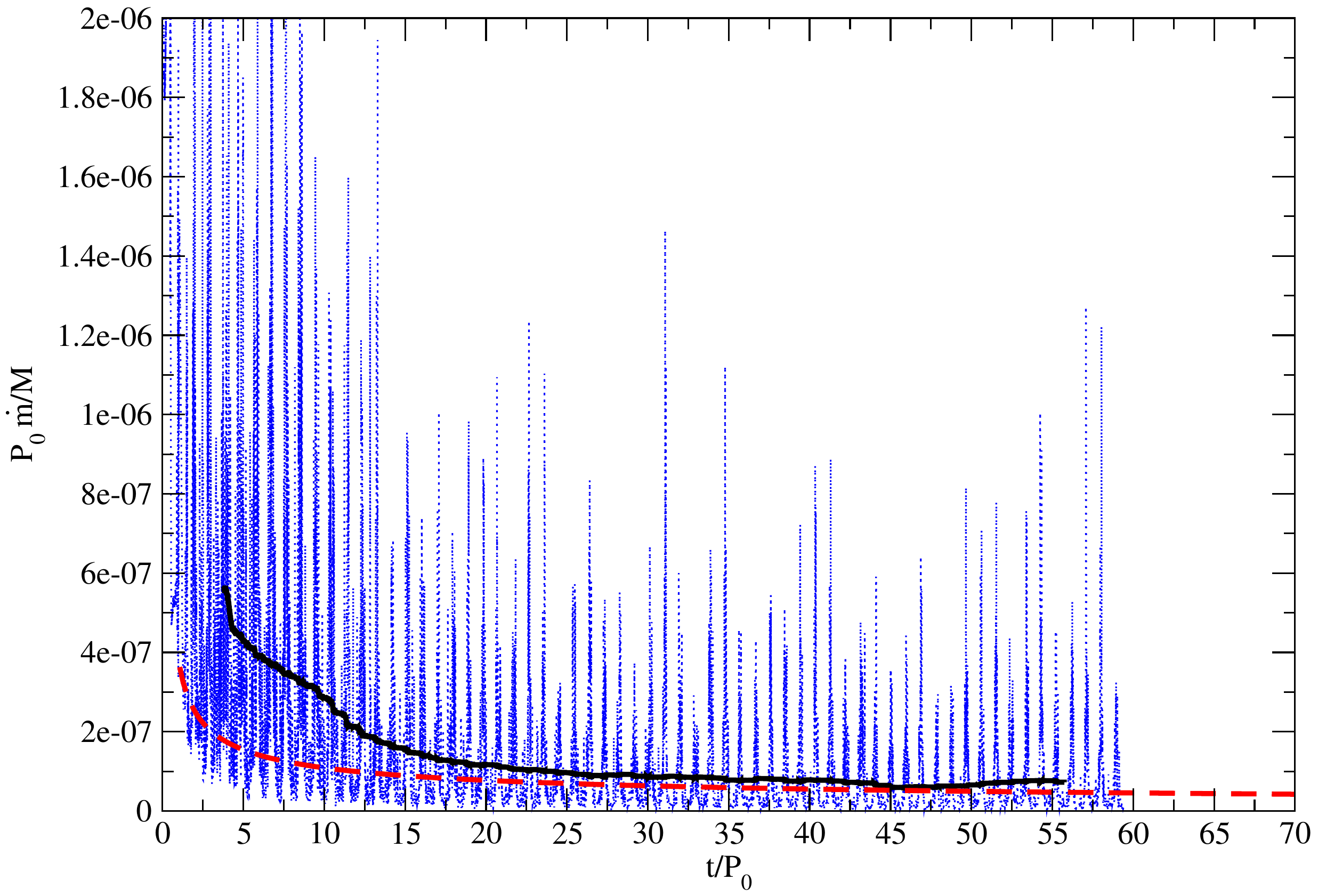}}
\caption{ The accretion rate onto the secondary in units $M/P_0$ as a function 
of time. The dotted  curve, showing large variations,  is obtained numerically, the solid curve represents 
the numerical results  time averaged over 7.5 periods, while the dashed  curve represents ${\dot m}_2$ as specified
through one of the  equations (\ref{e24}).} \label{Fig3n}
\end{figure}
 We consider the run with $q=2\cdot 10^{-2}$
and $\delta =0.05$ described in section \ref{Sims}. 
A  comparison  of the accretion rate measured in the  numerical simulation with our  analytical estimate is shown  in Fig. \ref{Fig3n}.
 Since in this case $r_a < h$ we use the second expression in (\ref{e24}) to make  the 
comparison. 
The  numerical data  ( dotted curve) shows sharp variations of the accretion rate.   These variations are due to the secondary approaching the  inner disk edge 
where the surface density is larger. Such approaches can occur because the inner disk edge is not circular and they repeat approximately periodically
 on  the  orbital time scale. This effect may have important observational consequences since it may lead to variation of the secondary luminosity.

  When the numerical data  is averaged over  time intervals of 7.5 orbital periods there is a good agreement between both approaches with the analytical curve giving somewhat
smaller values of $\dot m$. This may be  due to the fact that our analytical expression for $\Sigma_{min}$ underestimates
the minimum values of the azimuthally averaged  surface density in the gap that is found in simulations. 
 But note that  the accretion rate can reach values  an  order of magnitude
larger than the locally  averaged value.

\subsection{ A comparison of semi-analytic models and numerical simulations of migration in  an accretion disk of  finite extent in which a gap forms}
Here we compare  orbital  migration  determined using  equations (\ref{en1}), 
(\ref{en2}) and  (\ref{e14}) together with  (\ref{e15}) (see Section \ref  {orbtim}) with results of numerical simulations.
We performed  several two dimensional numerical simulations of accretion disks of 
 finite radial extent, $0.25 \le r/r_0  \le 5$.  We remark that for these disks the dependence of $\dot M$ on $t$  differs  from
that appropriate to the case of a  disk  of  formally infinite extent.
In this case, a  slightly different calculation of the dependence of
$\dot M(t)$ needs to  be undertaken to compare with the orbital evolution obtained from   simulations.

Here we  make a comparison  for a  sufficiently large mass ratio, $q=10^{-2},$ and two disk models. 
Model 1 employed  the non-linear viscosity law with $a=2/3$, 
$b=1$ and the initial surface density profile corresponding to zero angular momentum flux through the disk , i. e. given
by equation (\ref{en9}) with $F_0=1. $ Model 2  
adopted  a  constant  viscosity  for which  $a=b=0,$ and the initial surface density profile as for Model 1.
In both cases the dimensionless viscosity coefficient $\nu_0=10^{-5},$  $\delta = 0.05,$ we adopted standard softening  and the initial disk mass, $M_d=3\times10^{-3}M.$

\begin{figure}
\resizebox{\hsize}{!}{\includegraphics{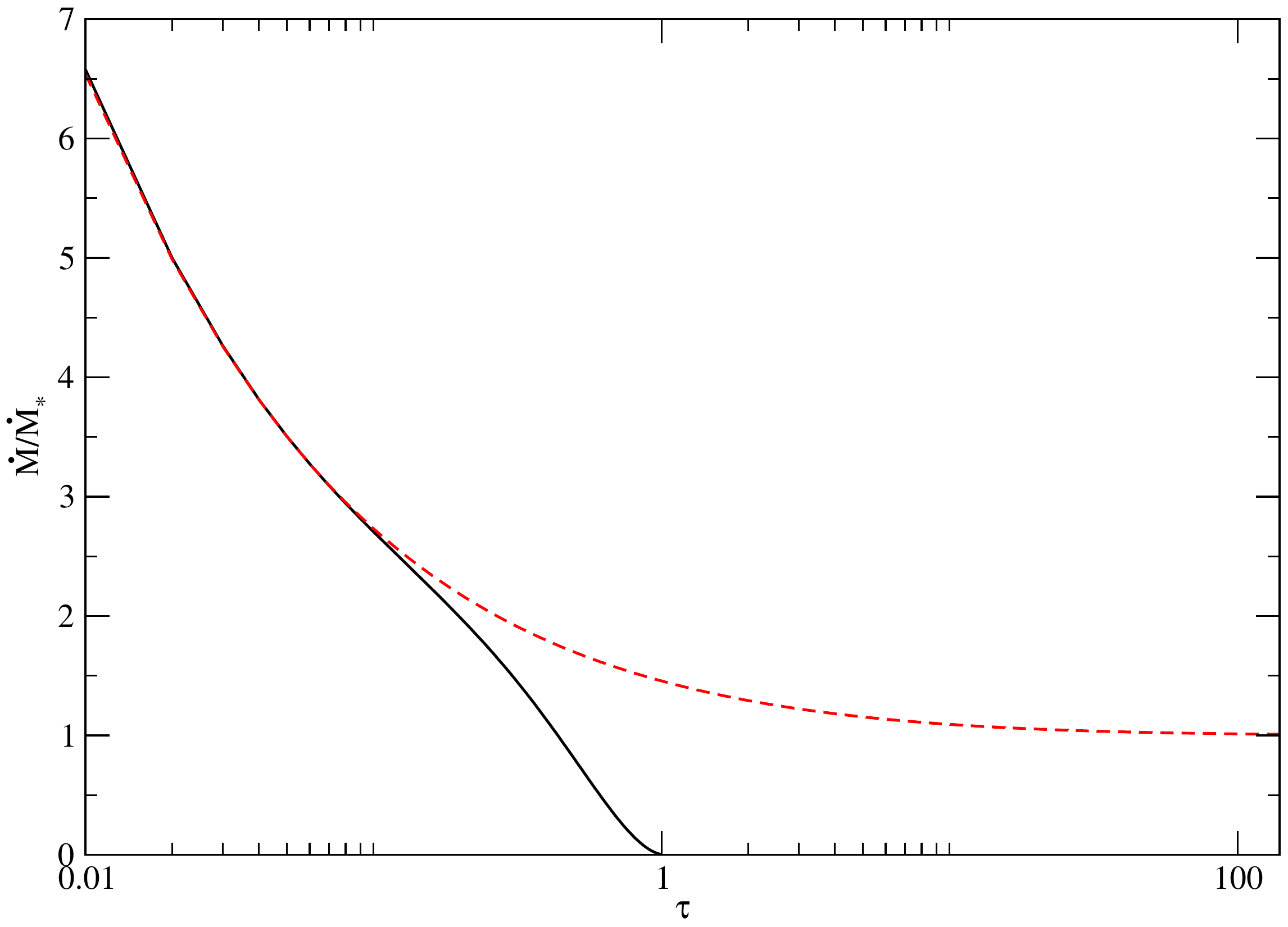}}
\caption{The dimensionless mass flux $\dot M/\dot M_*$ is shown as a function of time $\tau$ for the case
$a=2/3$, $b=1$ (model 1). The solid curve corresponds to our model of the disk of finite size. 
The dashed curve shows, for comparison,  the same dependence for the disk of formally infinite extent. 
} \label{mdot1}
\end{figure}

\begin{figure}
\resizebox{\hsize}{!}{\includegraphics{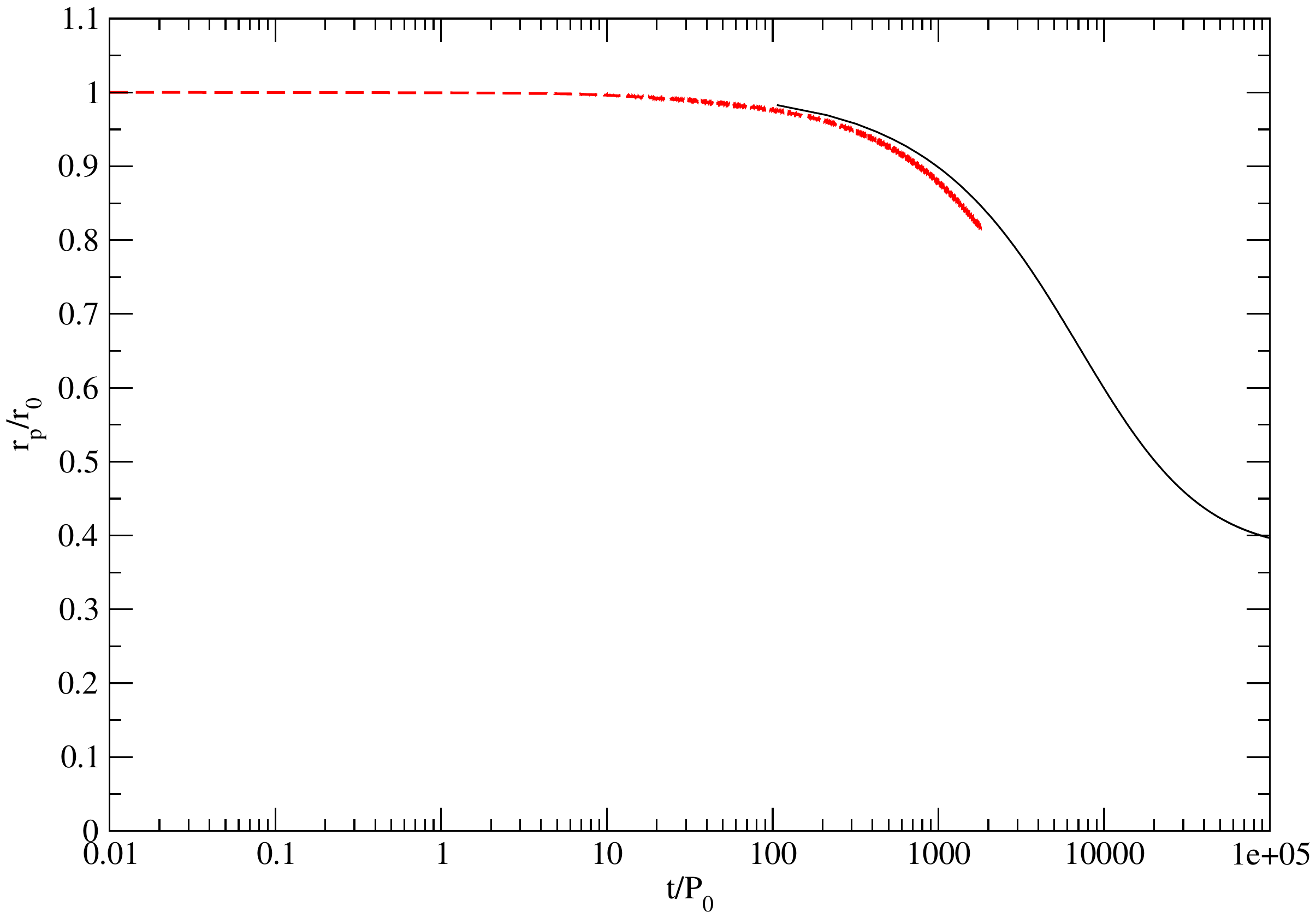}}
\caption{The dependence of the binary separation distance, $r_p/r_0$, as a function of $t/P_0$ for model 2. We recall 
that $r_0$ and $P_0$ are the binary separation distance and orbital period at the initial time $t=0$. The solid curve
is obtained from the  solution of equation (\ref{e15}) while the dashed curve represents the results of a two dimensional numerical simulation.} \label{at_nl}
\end{figure}

For  model 1 we solve equation (\ref{e15}) numerically, and use equations (\ref{en1}), (\ref{en2}), (\ref{e14})  and the definition 
of $\dot M_*$ to find the dependence of $r_p$ on time. The dimensionless mass flux for this model is shown in 
Fig. \ref{mdot1} as a function of time $\tau $ and the binary separation distance as a function of $t/P_0$ is shown
in Fig. \ref{at_nl} together with the results  obtained from  the  two dimensional numerical simulation.  
One can see that both approaches give  excellent agreement. 
The result based on the solution of (\ref{e15}) also predicts that eventually 
the evolution stalls, since on account of its being finite, all the disk mass is eventually accreted by the primary in this model. 

\begin{figure}
\resizebox{\hsize}{!}{\includegraphics{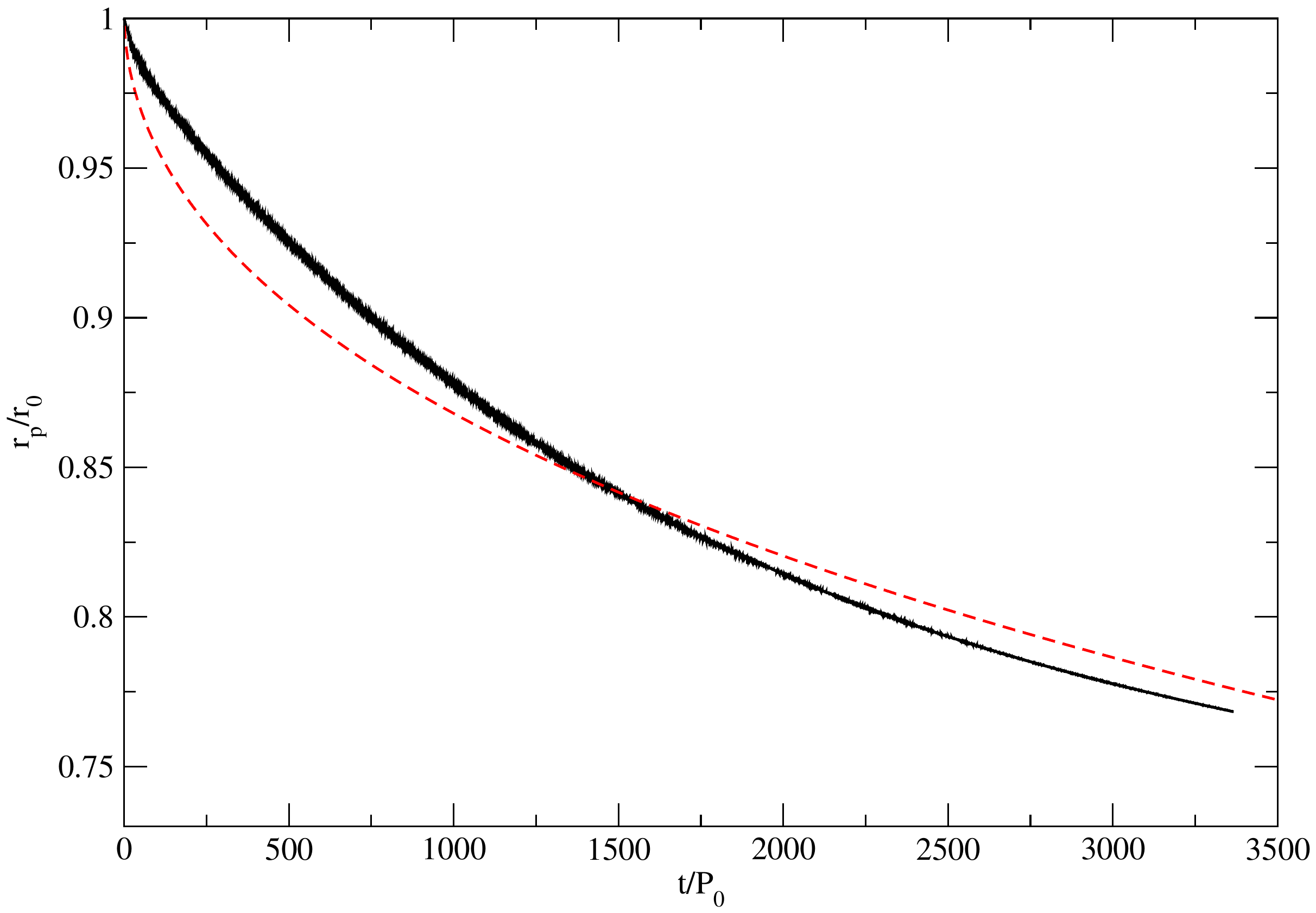}}
\caption{The dependence of the binary separation
distance on time. The solid curve shows the result of numerical
simulation with the mass ratio $q=10^{-2}$ and other parameters
described in the text,  the dashed curve is calculated
from  equation (\ref{ea28}).} \label{Fig3}
\end{figure}

For  model 2  we develop an analytic approach to the calculation of the evolution of $r_p$ in Appendix 
\ref{disc_ev}.  The result is given by equation (\ref{ea28}) and is compared with a two dimensional simulation
in Fig. \ref{Fig3}. As  for the non-linear case we have again a very good agreement between the numerical and
analytical approaches.

\section{Discussion}\label{Disc}
In this Paper we have studied the  evolution of a secondary mass embedded in an accretion disk
in a  quasi-circular  retrograde  orbit. Before summarising our conclusions we  briefly discuss some 
additional phenomena and processes that may be important under some circumstances.
We  consider   effects that could occur if the binary orbit possessed a moderate to large eccentricity
 and the importance of  orbital evolution  arising from the radiation of gravitational waves for perturbers
 orbiting close to the central mass.
\section*{Additional phenomena and processes}

\subsection{A binary with  large orbital  eccentricity}\label{eccorb}

A binary black hole may have a large eccentricity at the stage when interaction with the disk becomes important, (see 
e.g. Polnarev $\&$ Rees 1994). In other potential  astrophysical applications  of the scenario considered in the Paper,
for example   to  exoplanetary systems,  a massive planet can, in principal, have its  direction of orbital motion opposite to that
of the protoplanetary disk,  only as a result of its gravitation interaction with  other objects.  In this case, 
large eccentricities would  also be  expected.  Perhaps, the most important qualitative difference between the case of an 
eccentric binary and the case of circular binary considered  in this  Paper,  is that in the former case,  outer 
Lindblad resonances are still possible, though with  amplitudes  significantly reduced in comparison 
to the prograde case. Contrary to the impulsive and non-resonant torques considered in this Paper the Lindblad
torques are necessarily positive with respect to the direction of gas motion,  see e.g.  Goldreich $\&$ Tremaine  (1979).
Thus, if a disk is sufficiently thin,  there is a  possibility of gap formation in the 'standard 
sense'   with the disk gas being repelled  from  the binary orbit radii as the result of   an outward flux of positive
angular momentum supplied by waves launched at the resonances.   Let us briefly consider this possibility and estimate 
typical disk  parameters  for which a  gap could  be opened,  assuming that the binary has an eccentricity $e\sim 1$.

We use the standard expressions for the  resonant torques, $T_{ml}$,  associated with perturbations with  a particular azimuthal  mode number,
$m,$  and a particular time harmonic number $l$ using  expressions from  Artymowicz $\&$ Lubow  (1994),  hereafter AL, 
for the corresponding  component  of the gravitational potential due to  the perturber
  assuming  that the binary mean motion, or orbital  angular velocity, $\omega,$  entering these expressions is negative.
   Then,  terms with negative values of $l$ correspond to  waves propagating in the direction of the disk's rotation.
    It is readily found  (see e.g. AL) that the torques 
$T_{ml}$ scale approximately as $e^{2|m-l|}$ in the limit of small
eccentricities. Therefore, we parametrise them through the expression
\begin{equation}
T_{ml}=\tau_{ml}q^{2}e^{2|m-l|}\Sigma (GMa),
\label{d1}
\end{equation}
where $a$ is the semi-major axis which is clearly such that $a=r_p$ when $e=0.$ 

\begin{figure}
\resizebox{\hsize}{!}{\includegraphics{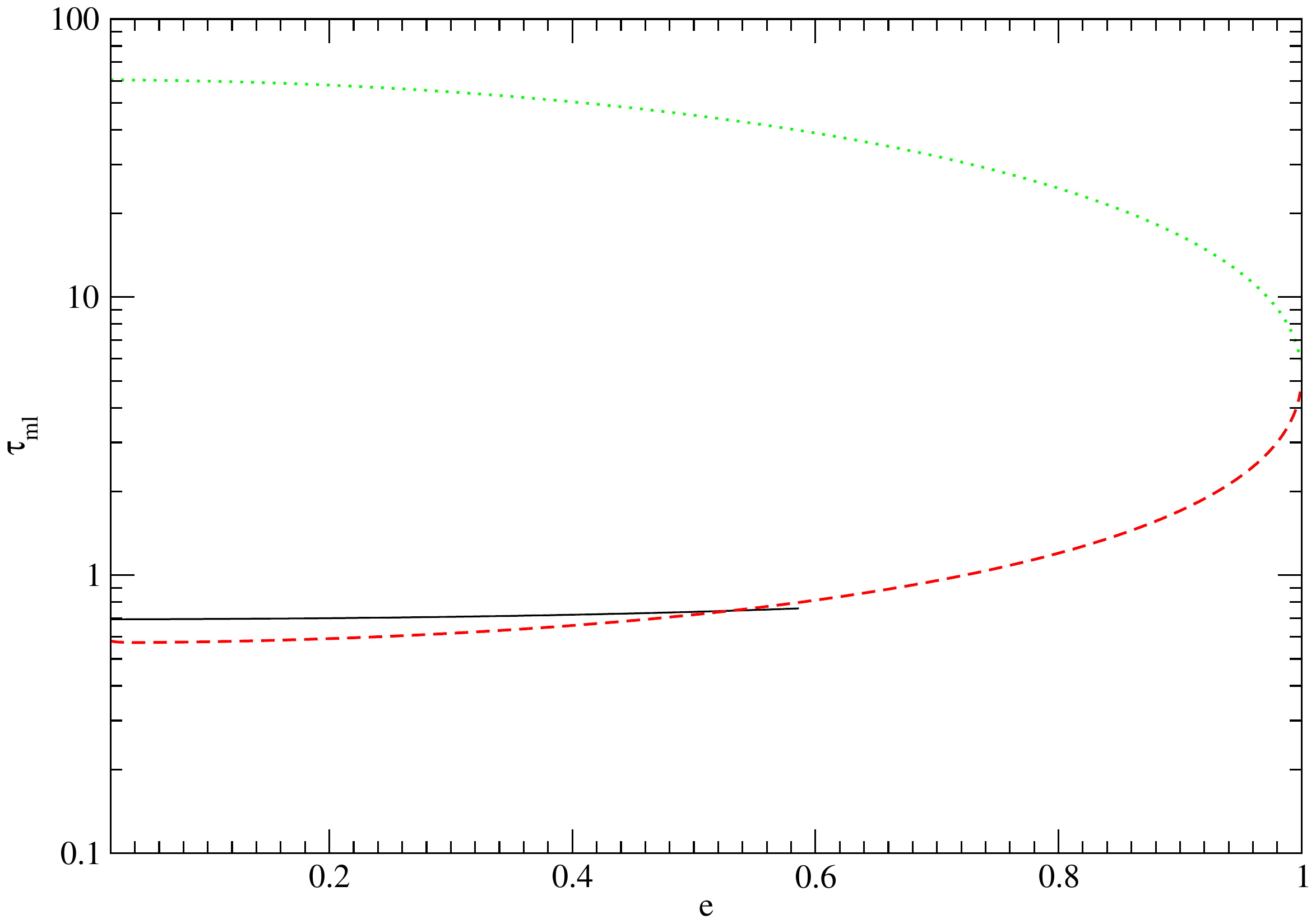}}
\caption{The dimensionless torques $\tau_{ml}$ are
shown for the cases $m=1,l=-1$ (solid curve),  $m=2,l=-1$ (dashed
curve) and  $m=2,l=1$ (dotted curve) as functions of eccentricity
$e$. Note that the solid curve ends at $e=0.58$. At this value of
eccentricity,  the apocentre distance $r_p=a(1+e)$ is equal to
$r_{(1,-1)}$ given by equation (\ref{d3}). 
At larger values of $e$ the corresponding resonance lies
within the binary orbit.  Also note that the case  $m=2,l=1$ shown for illustration
corresponds to prograde orbital motion of the binary. } \label{Disc1}
\end{figure}

The most important torques $T_{ml}$ are those corresponding 
to the smallest difference $|m-l|$  with $l < 0.$,  i.e.  terms  with $m=1$ and $l=-1$,   $m=2$ and $l=-1$ and so on. 
Assuming that $e$ is small one can use  expressions 
in AL to obtain
\begin{equation}
\tau_{(1,-1)}\approx 0.69, \quad \tau_{(2,-1)}\approx 0.57,
\label{d2}
\end{equation}
and the corresponding resonances
reside at radii
\begin{equation}
r_{(1,-1)}=4^{1/3}a, \quad r_{(2,-1)}=3^{2/3}a.
\label{d3}
\end{equation} 
A general dependence of these quantities on $e$ can be obtained numerically. This is shown in Fig. \ref{Disc1} together
with a  corresponding calculation  with  $m=2, l=1$ evaluated for a prograde orbit, which is shown for  comparison.
 
As indicated by  Fig. \ref{Disc1} the asymptotic expressions (\ref{d2}) are
approximately valid even in case of rather large eccentricities. In
particular, the numerical factor in the $(1,-1)$ term changes from
$0.69$ to $0.76$ when $e$ grows from zero to its maximal value
$\approx  0.58$ corresponding to the position of the resonance situated
at the apocentre distance. Analogously, the numerical factor of  the 
$(2,-1)$ term is smaller than unity for eccentricities smaller than
$0.72$. Thus, the expressions (\ref{d2}) can be used for estimates of
the corresponding torques at moderate  values of eccentricity. 
Accordingly  we  use  expressions for our estimates of conditions for  gap formation
below. 

In order to proceed  we use the criterion proposed in Lin $\&$ Papaloizou (1979b), see also AL. Namely,
we assume that a  gap  can be opened provided the angular momentum flux due to viscous forces,
$\dot L_*=3\pi \Sigma \nu \Omega r^2 < T_{ml}$. Note that $\dot L_*$ is evaluated at $r=r_{(m,l)}$.
Using this condition, equations  (\ref{en8add1}),   (\ref{d1}),   
and assuming that $\alpha $ and $\delta $ are small, 
we obtain  a  condition for  gap  formation in the form
\begin{equation}
e > e^{ml}_{crit}=\left( {3\pi \alpha \delta^2 r_{m,l} \over \tau_{ml} q^2 a}\right)^{{1/ (2|m-l|)}}.
\label{d5}
\end{equation} 
Now we use (\ref{d2}) and (\ref{d3}) to obtain
\begin{equation}
e^{1,-1}_{crit}\approx 0.2\alpha_*^{1/4}q_*^{-1/2}\delta_*^{1/2}
\label{d6}
\end{equation}
and 
\begin{equation}
e^{2,-1}_{crit}=0.37\alpha_*^{1/6}q_*^{-1/3}\delta_*^{1/3}
\label{d7}
\end{equation} 
for $m=1$, $l=-1$ and $m=2$, $l=-1$, respectively, where $\alpha_{*}=\alpha/10^{-2}$,  $q_{*}=q/10^{-2}$ and $\delta_*=\delta/10^{-3}$.

Since the critical eccentricities are of the order of $0.2-0.4$ for very thin accretion disks, which may be 
present in galactic nuclei, this effect may operate there. The situation is less favourable for protoplanetary disks,
where we typically have $\delta \sim 0.05 $, and, accordingly, $\delta_*\sim 50$. In this case we have 
the critical eccentricities formally exceeding unity for $\alpha_*=1$, and, therefore, this effect is unlikely to operate 
unless  $\alpha $ is very small. 

We comment that the gap opening process considered here differs from that considered in previous sections for near circular orbits
for which the disk edges are relatively close to the perturber.  For the case of a perturber in an eccentric 
orbit considered here, the gap is maintained by interactions at Lindblad resonances
such that the outer disk edge  may be  a distance of order $a$ even from apocentre. 
 Angular momentum transfer to the perturber  from material streaming through the cavity
due to viscous evolution of the disk may  occur  and also contribute to the maintenance of the gap. 
 However,  on account of a gap being produced through their action, 
  Lindblad torques are at least comparable to viscous torques.     
Accordingly  angular momentum exchange with the perturber due to Lindblad torques  and streaming material are potentially comparable.

If the gap opening by  Lindblad resonances can  indeed be
realised  and  these are important  for  the angular momentum exchange, the binary evolution would be quite different from that  described in this Paper. It may be closer 
to the regime discussed by IPP. Note, however, that non-resonant effects may 
interfere with  gap formation due to  resonances when apocentre approaches the  radius of a resonance.

 Also, note that when  resonances give the main contribution 
in the energy and angular momentum exchange between the binary and the disk,  the binary eccentricity decreases with 
time, and, therefore, the gap can be eventually overrun  by the disk gas, after which,  
the evolution will proceed in a manner similar to that  discussed in this  Paper\footnote
{Let us also remark that SPH simulations with a large mass ratio, $q=1$, indicate that when the eccentricity is sufficiently large, the inclination angle between the binary and
the disk may grow with time, see Roedig $\&$ Sesana (2014).}. The effects   discussed above are in need of  further investigation.

In this context we comment that the evolution of massive binary immersed in a circumbinary disk is also potentially  interesting in the context of
disk-planet interactions  (see e.g. Papaloizou et al. 2007) for a review.
 Although it is generally accepted that the planets are
formed in the protoplanetary disk rotating in the prograde sense, a gravitational interaction between a group of
planets (see e.g. Rasio $\&$ Ford 1996, Papaloizou $\&$ Terquem 2001) may, in principle, lead to the formation of
a planetary orbit of high eccentricity with the direction of orbital rotation opposite to that of a  disk of finite extent  and which enters into  it.
 Physically, this may be achieved by different generalisations of the well-known Lidov-Kozai effect, see e.g. Ziglin (1975), Farago $\&$ Laskar
(2010), Katz, Subo $\&$ Malhotra (2011) and  Li et al. (2013).

\subsection{The influence of emission of gravitational waves on the orbital evolution and accretion rate 
for  SBBH}\label{gravW}

In the case of SBBH there is an additional important mechanism  for driving  orbital evolution through emission of gravitational waves.
For a circular orbit and $q\ll 1$, the corresponding time scale can be easily obtained from  expressions given by
e.g. Landau $\&$ Lifshitz (1975) as
\begin{equation}
t_{gw}=\frac{5r_g}{8cq}\left(\frac{r_p}{r_g}\right)^4,
\label{w1}
\end{equation}
where $r_{q}=2GM/c^2 \approx 3\cdot 10^{13}M_{8}cm$, with $M_{8}=M/10^8M_{\odot}$, and $c$ denoting the speed of light from now on.

From equation (\ref{e21}) it follows that the time scale for  orbital evolution  due to interaction with the disk  can
be written in the form 
\begin{equation}
t_{ev}\approx 5\cdot 10^7 \left(\frac{ q_{-2} M_8}{\dot M_{-2} }\right) yr,
\label{w2}
\end{equation}
where $q_{-2}=q/10^{-2}$ and $\dot M_{-2}=\dot M/(10^{-2}M_{\odot}yr^{-1})$.
 From the condition $t_{gw} < t_{ev}$ we conclude that
gravitational waves  will determine the orbital evolution when
\begin{equation}
r_p< r_{gw(I)}= r_g\left(\frac{8cq t_{ev}}{5r_g}\right)^{1/4}\hspace{2mm} \approx \hspace{2mm}  \frac{0.7q_{-2}M_8}{ ({\dot M_{-2}})^{1/4}}\hspace{2mm}  pc.
\label{w3}
\end{equation} 
Note that the orbital period at $r_p \sim r_{gw(I)}$ being  given by     $P_{orb}\approx 5r_{-2}^{3/2}M_{8}^{-1/2}yr$,
where $r_{-2}=r_p/(10^{-2}pc)$ is expected to be of the order of a few years. 
From the definition of  $r_{gw(I)}$ and (\ref{w1}) it also follows that
\begin{equation}
t_{gw}=\left({r_p\over r_{gw(I)}}\right)^{4}t_{ev}. 
\label{w4}
\end{equation}

Another important scale, $r_{gw(\nu)}$, is determined by the condition  that the time scale for orbital evolution
due to gravitational radiation be less than the time scale for viscous evolution of the disk, or  $t_{gw}(r_{p} < r_{gw(\nu)}) < t_{\nu}.$
 Making  use of  (\ref{en8add1}) we obtain 
\begin{equation}
r_{gw(\nu)}=r_g \left[\frac{32\sqrt{2}q}{15\alpha\delta^2 }\right]^{2/5} \approx 
5\cdot 10^{-3}M_{8}(q_{-2})^{2/5}\alpha_{*}^{-2/5}
\delta_{*}^{-4/5} pc.
\label{w5}
\end{equation}
When $r < r_{gw(\nu)}$ the absolute value of the radial velocity of the perturber exceeds that of the disk
gas. In this regime the disk gas is transferred from the inner part of the disk  to the outer part opposite 
to the regime considered above. However, arguments leading to the expression (\ref{e24n}) remain essentially 
the same if instead of the accretion rate through the disk, $\dot M$, the rate of transfer of the disk gas
through perturber's orbit, $\dot M_{tr}$, is used. Note that $\dot M_{tr}$ is defined in the frame, where 
perturber is at rest. We can estimate it as $\dot M_{tr}\sim M_{d}(r < r_p)/t_{gw}$, where the disk mass inside 
the perturber's orbit can be found from the first expression of  (\ref{e13}).
 On the other hand, as  discussed
above the disk inside the perturber's orbit may be approximated as a stationary disk  characterised by the accretion
rate $\dot M$, and, therefore, its mass can be estimated as $M_{d}(r < r_p)\sim \dot Mt_{\nu}$. Taking these considerations
into account we obtain
\begin{equation}
\dot m \sim \frac{q^{1/3} M_{d}(r < r_p)}{t_{gw}}\sim \frac{q^{1/3}\dot M t_{\nu}}{ t_{gw}} \sim\frac{ q^{1/3}\dot M r^4_{gw(\nu)}}{ r^{4}}.
\label{w6}
\end{equation}
This indicates that the  accretion rate onto the secondary can exceed that onto the primary,  $\sim \dot M,$ provided that 
\begin{equation}
r < r_{crit}=q^{1/12}r_{gw(\nu)}.
\label{w7}
\end{equation}
Since the power of $q$ in (\ref{w7}) is small,   we have  that typically $r_{crit}\sim r_{gw(\nu)}$.  Note, however, that 
even when $r < r_{crit}$ the  luminosity of the secondary does not necessarily exceed that of the primary. 
This is also determined
whether the mode of accretion onto the secondary is disk-like or more  approximately  spherical. This issue  is one for  future  investigation.

\subsection{Secular evolution of directions of angular momenta of the binary and the disk}\label{twd}

As  mentioned in Introduction the gravitational torque exerted by  a stationary twisted circumbinary  disk on the binary orbit  changes its
orientation on a long timescale $t_{or}$.  It is assumed that the disk is aligned with the binary orbital plane
at scales smaller than the alignement radius $r_{al}$, but misaligned at scales $r_{al}$ with some inclination angle, which may be initially small but  subsequently may become
large enough for non-linear effects to become important.
The action of the disk  torque  tends  to overturn the binary orbit when it is retrograde, see e.g. IPP. 

When the inclination angle is large enough  non-linear effects are important,  and for an estimate of the time $t_{or},$  we formally assume that the effective $\alpha $ for the twisted disk
$\sim 1,$
see e.g. Ogilvie (1999) for a theoretical justification and also Lodato $\&$ Price (2010) for numerical SPH simulations 
of the evolution of twisted disks in the regime of large gradients of the disk's tilt.
\footnote{Note that there are non linear  effects operating in twisted disks with  finite   inclination angles, which are not considered
in Ogilvie (1999), such as a possibility of standing shocks, see e.g. Fragile $\&$ Blaes (2008). These effects could lead to additional damping of 
non-stationary twisted disturbances once the inclination to the binary orbit exceeds $\sim \delta,$  see e.g. Sorathia et al.  (2013),  and may act so as to produce  an increase of the  effective $\alpha$. Also, the presence of 
shocks may increase the disk thickness since they  provide an additional source of energy dissipation.}
We then use the simple estimate of
$t_{or}$ given in IPP as 
\begin{equation}
 t_{or}\sim \sqrt{{r_p\over r_{al}}}{M_p\over \dot M}, 
\end{equation}
 where the alignement radius $r_{al}\sim r_{p} \sqrt{q}/ \delta $
 (see  equation (23) of IPP).
  Recalling that the orbital evolution timescale $t_{ev}=M_p/( 2\dot M)$ we get
\begin{equation}
t_{or}\sim 0.2\delta_*^{1/2}q_*^{-1/4}t_{ev}.
\label{tor}  
\end{equation}
Note that although the numerical factor in (\ref{tor}) is clearly approximate,  it would increase if a smaller value of the effective $\alpha$ had been  used as long as  it exceeded $\delta.$
 One sees that as long as $r_{al} \sim r_p,$ as is the case for the gap forming simulations discussed in Section
 \ref{Sims},   $t_{or}\sim t_{ev},$  which occurs because the process inducing inclination changes works at the same rate as that producing changes to the semi-major axis.
 In that case changes in the orientation of the orbit may not be significant.
 On the other hand, for some values of $q$ and $\delta,$ $r_{al}$ could exceed $r_p$ significantly enough
that the timescale for  the orientation evolution could be somewhat smaller than the orbital evolution timescale leading to significant changes in inclination.
 This is an aspect  for future study.

\section{Summary of results and conclusions}\label{sumconc}

In this Paper we have considered  a binary  with   small mass ratio $q\ll 1$ embedded in an accretion disk with the direction of orbital motion
being retrograde with respect to the  rotation of the disk. 
We studied the evolution of the semi-major axis  and the mass accretion rate onto the secondary component  focusing  on  the case of a quasi- circular orbit.

We  employed several approaches to the  problem.   A simple semi-analytic approach based on
solving a one dimensional diffusion equation for the disk surface density  that incorporated  a local scattering  model for estimating the rate of angular momentum
exchange between the orbit and disk, which resulted in evolution of the orbit. 
In addition to this we presented a  simplified   purely analytic calculation of the surface density profile
in the vicinity of the perturber and the construction of a similarity solution for the outer disk that was applicable when the latter was of infinite extent.
Under the assumption of slow orbit evolution, valid when the secondary mass is much larger than the local disk mass, the rate of orbital evolution
in the form of an inward migration could be readily calculated.  In addition to these  approaches
 we performed two-dimensional numerical  hydrodynamical simulations.  All methods 
were  found to be in qualitative, and often, as in the context of orbital semi-major axis  evolution,  in quantitative agreement.

We found that  for binaries with very small  mass ratios, the secondary component  migrated through the disk leaving the surface density   virtually unchanged.
This corresponds to the usual type I migration regime for prograde orbits. The orbital evolution is due
 to the launching of tightly wound  density  waves that  propagate  away from the orbit (see section \ref{s2}).
We calculate the evolution rate, which is much less than in the prograde case,

When the mass ratio is sufficiently large,  $q > \sim 1.57\delta^2,$  $\sim 10^{-3}-10^{-2},$ for typical parameters considered, a gap in the disk opens in the vicinity of the perturber's orbit 
(see sections \ref{Gapf} and \ref{Sims}).
However, the form of the  gap  and the way it opens  are not the same as in the prograde case.
 In the retrograde case gap opening  is produced by efficient  removal of angular momentum in the vicinity of the orbit as the disk material passes through it.
 This leads  to a depression of the  surface density profile near the orbit that we are able to match  quite well  using  simplified  analytic modelling. 

Provided that the secondary  mass is significantly larger than the  typical disk  mass in the vicinity of   its orbit,  the disk exterior to the gap over a length scale of order 
$r_p$ relaxes to a quasi-stationary state characterised by the rate of mass flow through the disc, $\dot M$,  on a time scale  that is  characteristic of the
local viscous time scale,  but significantly  shorter  than the characteristic
time scale of orbital evolution, $t_{ev}.$ 
  For a disk of large  radial extent $\dot M$ is comparable to  the accretion rate at large distances. 
  The orbital evolution time scale, $t_{ev},$  was found to be 
of $M_p/(2\dot M)$, where $M_p$ is the secondary mass (see section \ref{orbtim}).

When the orbital evolution is solely determined by interaction with the disk, the time averaged  accretion
rate onto the secondary, $\dot m$, is estimated to be at most 
$\dot m \sim q^{1/3}\dot M,$ being  significantly  smaller than  the accretion rate onto the primary $\sim \dot M$ (see section \ref{est}).  This behaviour is
different from what is seen in the prograde  case,  where the accretion rates can be  reduced due to the presence of a deep cavity in the disk  on  scales $\sim r_p.$ Our numerical results also show  large variations of the accretion 
rate on the orbital timescale.
If the radiative efficiency of  the accretion onto the secondary  is sufficiently large,  the binary may manifest itself as a non-stationary source of radiation with
typical time scale for  variability   on the order of  the orbital period.  It  may also  be observed as two nearby sources of radiation being  blueshifted and  redshifted with
respect to each other. 

In the case of a  supermassive binary black hole the orbital evolution is also significantly influenced by emission of gravitational waves. This effect becomes more
important than the interaction with the disk  when $r_p  < 10^{-2}pc$, for typical parameters of the problem. When the evolution is governed by  gravitational wave
emission such that its inward radial drift  speed exceeds the gas radial velocity  induced by disk viscosity,  it is expected that  the accretion rate onto the perturber would increase,
  possibly becoming  larger than $\dot M$ for  a brief period of time (see section \ref{gravW}). 
Another  open issue for   future study is  a behaviour of the system when the orbit has  an appreciable eccentricity.
 We indicated in section \ref{eccorb} that when the binary is very eccentric 
and the disc is very thin,  a conventional gap may be opened due to the effect  Lindblad resonances,  which are present in this case even for  binaries with retrograde orbital motion
relative to the disk.  In addition to   a systematic detailed  study of effects arising from orbital  eccentricity, 
  another  issue  for future  consideration,   is the influence of  a mutual inclination of the binary orbital plane with respect to the disk mid plane and
the effect of secular evolution of the binary inclination with respect to the disk plane at large distances.

\section*{Acknowledgements}

We are grateful to R. R. Rafikov, V. V. Sidorenko and the referee for useful comments.

\noindent  PBI was supported in part by programme 22 of  the Russian Academy of Sciences and in part by
the Grant of the President of the Russian Federation for Support of Leading
Scientific Schools of the Russian Federation NSh-4235.2014.2.

{}

\begin{appendix}

\section{The evolution of a finite  disk  with  $a=b=0$} 
\label{disc_ev}
When $a=b=0$ equation (\ref{e15}) becomes  a
linear equation taking on  the  simple form
\begin{equation}
{\partial \tilde \Phi \over \partial \tau} ={1\over h^2}
{\partial^2 \tilde \Phi \over \partial h^2}, \label{ea16}
\end{equation}
where $\tilde \Phi = h\tilde \Sigma$. 
We express  the solution of
(\ref{ea16}) for  $h \in (1,h_{out})$  as a linear combination of normal modes in the form 
\begin{equation}
\tilde \Phi =\sum_i a_i \phi_i(h)e^{-\lambda_i\tau} . \label{ea17}
\end{equation}
The boundary conditions  satisfied by  $\Phi(h,\tau )$ ( and also $\phi_i(h), i = 1,2...$)  are 
\begin{equation}
\tilde \Phi(h=1)=0, \quad {\rm and}\quad  {\partial \Phi(h=h_{out})\over \partial h}
=0. \label{ea18}
\end{equation}
The required solution to (\ref{ea17}) should also satisfy the initial
condition $\Phi(\tau=0)=1.$

\begin{figure}
\resizebox{\hsize}{!}{\includegraphics{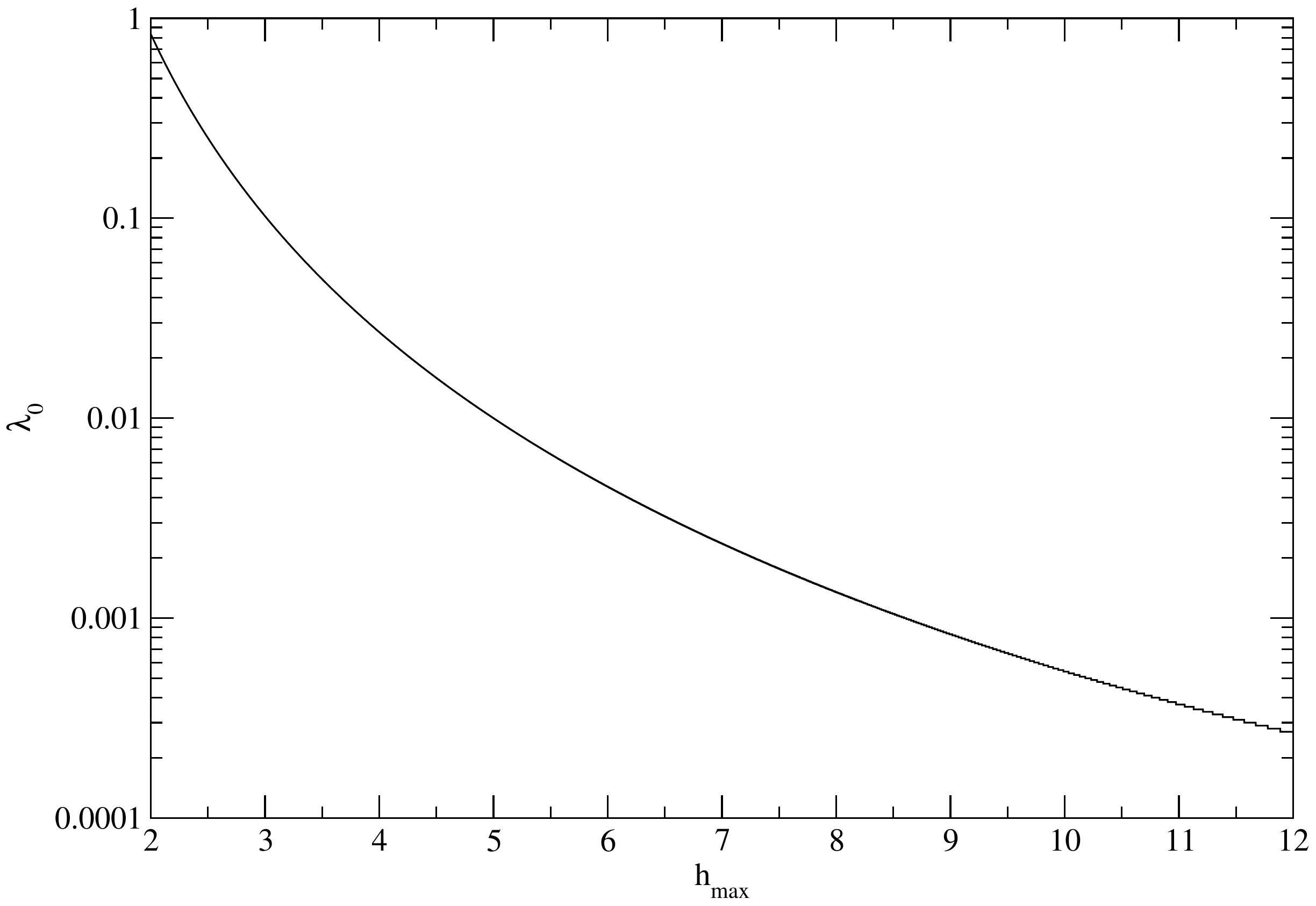}}
\caption{The minimum eigenvalue $\lambda_0 $ as a
function of the outer boundary  location  of the computational domain. }
\label{Fig2}
\end{figure}

 Let recall that in order to express
the surface density in physical units, we must multiply $\tilde
\Sigma$ by product of a numerical factor and $M/r_p^2$, where the
numerical factor is determined by the  specification of the ratio of the disk  mass to
the mass of the central body within the computational domain. For
the particular  numerical simulations  discussed in this Paper,
the  numerical factor relating the quantity
$\Phi=\Sigma/\sqrt{GMr_p}$ to $\tilde \Phi$, $\Phi_0$, is
approximately equal to $2.4\cdot 10^{-4}$. Note, however, that the
initial condition is incompatible with the inner boundary
condition (\ref{ea18}). This is  due to the fact that our
boundary conditions are strictly valid only after some period of time has
elapsed.  Nonetheless,  the 
choice of  the functions $\phi_i(h)$ enables
 the initial condition to be satisfied everywhere apart from  at $h=1,$  see below. After
substitution of (\ref{ea17}) to (\ref{ea16}) we get
\begin{equation}
{d^2  \phi_i \over d h^2} + \lambda_i h^2\phi_i=0.  \label{ea19}
\end{equation}
Note that the solutions to (\ref{ea19}) are orthogonal with respect
to the scalar product
\begin{equation}
<\phi_i |\phi_j>=\int^{h_{out}}_{1}dh h^2 \phi_i \phi_j.
\label{ea20}
\end{equation}
They can be expressed in terms of  Bessel functions as
\begin{equation}
\phi_i=\sqrt{h}(C_1 J_{1/4}(z)+ C_2 J_{-1/4}(z)), \label{ea21}
\end{equation}
where $z=\lambda_i h^2/2$. The boundary conditions (\ref{ea18})
then define an eigenvalue  problem leading to a discrete set of positive
$\lambda_i$ bounded from below. 
We arrange the eigenvalues in such a way that larger eigenvalues correspond to larger
values of $i$, and calculate them numerically for
$i=0, 1, .. 13$, with $i=0$ corresponding to the mode with the smallest
value of $\lambda $ for a given $h_{out}.$
The dependence
$\lambda_0$ on $h_{out}$ is shown in Fig. \ref{Fig2}.

When $\lambda_i \gg 1$ we can use the asymptotic representation of
the Bessel functions corresponding to large values of their
arguments. In this case it is easy to show that $\lambda_i$
approximately satisfy
\begin{equation}
\lambda_i\approx \pi^2\left ({1 + 2i\over h_{out}^2-1}\right )^2. \label{ea22}
\end{equation}
We have checked that the approximate values (\ref{ea22}) are close
to the ones obtained numerically even for $i=1.$

In order to find the coefficients $a_i$ in (\ref{ea17}) we use the
formal identity $1=\sum_i a_i \phi_i(h)$  and the orthogonality
condition (\ref{ea20}) to get
\begin{eqnarray}
a_i&=&{I_2\over I_1}, \quad {\rm with}\quad  I_1=\int_{1}^{h_{out}}dh h^2
\phi_i(h)^2\quad{\rm and} \nonumber\\
 I_2&=& \int_{1}^{h_{out}}dh h^2 \phi_i(h)=
{1\over \lambda_i}{d \over dh}\phi_{i}(h=1), \label{ea23}
\end{eqnarray}
where we have used  equation (\ref{ea19}) to obtain the last
equality.

For our purposes we need to know the value of the quantity  $F_-\equiv \partial \tilde \Phi/ \partial h, $
evaluated at $h=1.$ 
It is easy to see that this can be represented in the form
\begin{equation}
F_{-}=\sum_i \lambda_i \Psi_i e^{-\lambda_i \tau}, \quad {\rm with}\quad
\Psi_i={a_i\over \lambda_i}{d \over dh}\phi_{i}(h=1).
\label{ea24}
\end{equation}

In general the coefficients $\Psi_i$ have to be  be calculated numerically.
Thus  we obtain $\Psi_0=3.1$.
For large values of $i$ the  asymptotic representation of the Bessel
functions may be used,  as for eigenvalue  determination,   with the result
\begin{equation}
\Psi_i\approx {4\over \lambda_i (h_{out}^2-1)}, \label{ea25}
\end{equation}
where we can use equation (\ref{ea22}) to specify $\lambda_i.$  Note
that equation (\ref{ea25}) is found to give  reasonable values for
$\Psi_i$ even when $i$ is small.  Thus  for $i=1$  the
numerical value of  $\Psi_1\approx 0.2$, whereas  equation (\ref{ea25})
gives $\Psi_1\approx 0.18$.

\subsection{Orbital evolution}
Now let us consider the orbital evolution. 
In order to find the dependence
of $r_p$ on time we  again  use equations (\ref{en1}), (\ref{en2}) and (\ref{e14})
recalling that the relationship between the mass flux $\dot M$
and the quantity $F_-$ as given by 
 $\dot M = \dot M_* F_-=3\pi
\nu_0 \Sigma_0 F_-$ (see equation (\ref{en8}) and preceding discussion).
 We  take  account of the effect of 
evolution of perturber's orbital distance $r_p$ with time,  on the surface density profile,   in the
simplest possible manner, by  utilising an appropriate scaling of quantities
of interest with the ratio $r_p/r_0$ as in the main text.
We recall  that since the inner boundary
condition in (\ref{ea18}) assumes that the variable $h$ is always
equal to unity at the perturber's position it is convenient to
redefine this quantity according to   $h=\sqrt{r/r_p}$, which strictly
coincides with the older definition only when $t=0$. Furthermore, we
redefine the dimensionless time $\tau$ assuming that in equation
(\ref{e7})  $r_p$ replaces  $r_0.$ Thus  we have $\tau =3\nu_{0} t/ r_p^2 $. However,  the definition of the
dimensionless surface density $\tilde \Sigma $ and  the relation
between $\nu $ and $\Sigma $ given by equation (\ref{e5})  remain unchanged.

 We proceed by introducing  the dimensionless time
variable $\tilde t=t/P_0$, where $P_0=2\pi r_0^{3/2}/\sqrt{GM}$
and  the dimensionless orbital distance $\tilde r_p=r_p/r_0$.
Utilising  equations (\ref{en1}), (\ref{en2}) and (\ref{e14}), it is straightforward to obtain an
 equation governing  the orbital evolution in terms of these quantities.
This takes the form
\begin{equation}
{1\over \tilde r_p}{d  \tilde r_p \over d\tilde t}=-{12\pi^2\over
q}\nu_* \Sigma_* F_{-}, \label{ea26}
\end{equation}
where $\Sigma_*=r_0^2\Sigma_0/M$ and $\nu_*=\nu_0/\sqrt{GMr_0}.$
For the simulation discussed above,  $\Sigma_*\approx 2.4\cdot 10^{-4}$ and
$\nu_*=10^{-5}.$

From equation (\ref{ea24}) it follow that $F_-$ can be written in
the form
\begin{equation}
F_-=\sum_i \lambda_i \Psi_i e^{-\Lambda_i \tilde t} \quad {\rm with}\quad
\Lambda_i= {3\pi \over 2\tilde r_p^2}\nu_*\lambda_i . \label{ea27}
\end{equation}
We integrate (\ref{ea27}) formally assuming that $r_p$ does not
depend on time on the r.h.s. to obtain
\begin{equation}
\tilde r_p=e^{-\Delta }, \quad \Delta = {8\pi \tilde r_p^2\over
q}\Sigma_*\sum_i \Psi_i (1-e^{-\Lambda_i \tilde t}). \label{ea28}
\end{equation}
Note that equation (\ref{ea28}) provides an implicit
dependence of $\tilde r_p$ on time since $\tilde r_p$ enters both
sides of the equation.  Its numerical solution by a  standard iterative
procedure is shown in Fig. \ref{Fig3}.

\end{appendix}


\begin{thebibliography}{}

\bibitem[]{amar} 
Amaro-Seoane,  P.,  Aoudia, S.,  Babak, S. ,  et al. , 2013,  GWN, 6, 4

\bibitem[]{arm} Armitage, P. J., Natarajan, P., 2002, ApJ, 567L, 9

\bibitem[1994]{Art} Artymowicz, P., Lubow, S. H., 1994, ApJ, 421, 651

\bibitem[]{beg} Begelman, M. C., Blandford, R. D., Rees, M. J., 1980, Nature, 287, 307

\bibitem[]{i19} Bogdanovic, T., Smith, B. D., Sigurdsson, S., Eracleous, M., 2008, ApJS, 174, 455

\bibitem[]{i17} Bogdanovic, T., Eracleous, M., Sigurdsson, S., 2009, New Astronomy, 53, 113

\bibitem[1944]{Bon}Bondi, H.,  Hoyle, F., 1944, MNRAS, 104, 273

\bibitem[]{Burke} Burke-Spolaor, S.,  2013, CQGRA, 30, 4013

\bibitem[]{Corsini}
Corsini, E.M. 2014, in Counter-Rotation in Disk Galaxies, ASP
Conference Series, Vol. 486, Eds. E. Iodice \& E. M. Corsini
(ASP: San Francisco), p. 51

\bibitem[]{i18} Cuadra, J., Armitage, P. J., Alexander, R. D., Begelman, M. C., 2009, MNRAS, 393, 1423

\bibitem[]{i31} Farago, F., Laskar, J., 2010, MNRAS, 401, 1189

\bibitem[]{i12} Farris, B. D., Liu, Y. T., Shapiro, S. L., 2011, Phys. Rev. D., 84, 4024

\bibitem[]{i4} Farris, B. D., Duffell, P., MacFadyen, A. I., Haiman, Z., 2014, ApJ, 783, 73

\bibitem[]{frag} Fragile, P. C., Blaes, O. M., 2008, ApJ, 687, 757

\bibitem[]{gr} Grishchuk, L. P., Lipunov, V. M., Postnov, K. A., Prokhorov, M. E., Sathyaprakash, B. S.,
2001, Physics Uspekhi, 44, 1

\bibitem[]{i3} Gold, R., Paschalidis, V., Etienne, Z. B., Shapiro, S. L., Pfeiffer, H. P.,
2014, Phys. Rev. D., 89, 4060

\bibitem[1979]{Gold} Goldreich, P., Tremaine, S., 1979, ApJ, 233, 857

\bibitem[]{i26} Gould, A., Rix, H.-W., 2000, ApJ, 532L, 29

\bibitem[]{i16} Haiman, Kocsis, B., Menou, K., 2009, ApJ, 700, 1952 

\bibitem[]{i6} Hayasaki, K., Saito, H., Mineshige, S., 2013, PASJ, 65, 86

\bibitem[]{i36} Ivanov, P. B., Igumenshchev, I. V., Novikov, I. D., 1998, ApJ, 507, 131

\bibitem[1999]{Iva} Ivanov, P. B., Papaloizou, J. C. B., Polnarev, A. G., 1999, MNRAS, 307, 
79

\bibitem[]{i5} Ju, W., Greene, J. E., Rafikov, R. R., Bickerton, S. J., Badenes, C., 2013, ApJ, 777, 44

\bibitem[]{i33} Katz, B., Dong, S., Malhotra, R., 2010, Phys. Rev. Lett., 107, 1101

\bibitem[]{king} King, A. R., Lubow, S. H., Ogilvie, G. I., Pringle, J. E., 2005, MNRAS, 363, 49

\bibitem[]{i8} Kocsis, B., Haiman, Z., Loeb, A., 2012, MNRAS, 427, 2680

\bibitem[]{komb}  Komberg, B. V., 1968, Soviet Astronomy, 11, 727

\bibitem[]{i22} Komossa, S., 2006, Memorie della Societa Astronomica Italiana, 77, 733

\bibitem[]{land} Landau, L. D., Lifshitz, E. M., 1975, The Classical Theory of Fields. Vol. 2, 4th ed., Butterworth-Heinemann

\bibitem[]{i35} Li, G., Naoz, S., Kocsis, B., Loeb, A., 2013, ApJ, 785, 116

\bibitem[1979]{Lin} Lin, D. N. C., Papaloizou, J., 1979, MNRAS, 186, 799

\bibitem[1979]{Lin1} Lin, D. N. C., Papaloizou, J., 1979, MNRAS, 188, 191 (b)

\bibitem[1986]{Lin11} Lin, D. N. C., Papaloizou, J., 1986, ApJ, 309, 846

\bibitem[1993]{Lin2} Lin, D. N. C., Papaloizou, J. C. B., 1993, in:
Protostars and planets III (A93-42937 17-90), p. 749-835

\bibitem[]{i24} Liu, F. K., 2004, MNRAS, 347, 1357

\bibitem[]{lod} Lodato, G. Nayakshin, S., King, A. R., Pringle, J. E., MNRAS, 2009, 
398, 1392

\bibitem[]{lod1} Lodato, G., Price, D. J., MNRAS, 2010, 405, 1212

\bibitem[]{i21} Loeb, A., 2007, Phys. Rev. Lett., 99, 1103

\bibitem[]{i23} Lobanov, A. P., Roland, J., 2005, A$\&$A, 431, 831

\bibitem[]{lyn} Lynden-Bell, D., Pringle, J. E., 1974, MNRAS, 168, 603

\bibitem[1987]{Lyu} Lyubarskiy, Y. E., Shakura, N. I., 1987, Soviet Astronomy Letters, 13, 386

\bibitem[]{i20} MacFadyen, A. L., Milosavljevic, M., 2008, ApJ, 672, 83

\bibitem[]{i1} McKernan, B., Ford, K. E. S., Kocsis, B., Lyra, W., Winter, L. M., 2014, 
MNRAS, 441, 900

\bibitem[]{i13} Montuori, C., Dotti, M., Colpi, P., Decarli, R., Haart, F., 2011, MNRAS, 412, 26

\bibitem[]{i9} Montuori, C., Dotti, M,. Haardt, F., Colpi, M., Decarli, R., 2012, MNRAS, 425, 
1633

\bibitem[2011]{Nix} Nixon, C. J., King, A. R., Pringle, J. E., 2011, MNRAS, 417, L66 

\bibitem[2011]{Nix1} Nixon, C. J., Cossins, P. J., King, A. R., Pringle, J. E., 2011, MNRAS, 412, 1591 

\bibitem[1999]{Og} Ogilvie, G. I., 1999, MNRAS, 304, 557

\bibitem[2009]{Par}Paardekooper, S.-J.,  Papaloizou, J. C. B., 2009, MNRAS,  394, 2297

\bibitem[1984]{Lin3} Papaloizou, J., Lin, D. N. C., 1984, ApJ, 285, 818

\bibitem[]{i30} Papaloizou, J. C. B., Terquem, C., 2001, MNRAS, 325, 221

\bibitem[]{i30a} Papaloizou, J. C. B., Terquem, C., 2006, Reports on Progress in Physics, 69, 119

\bibitem[2012]{Pod} Podlewska-Gaca E., Papaloizou, J. C. B., Szuszkiewicz, E.,   2012, MNRAS, 421, 1736 

\bibitem[1994]{Pol} Polnarev A. G., Rees, M. J., 1994, A$\&$A, 283, 301

\bibitem[]{i7} Rafikov, R., 2013, ApJ, 774, 144

\bibitem[]{i29} Rasio, F. A., Ford, E. R., 1996, Science, 274, 954 

\bibitem[]{i25} Rieger, F. M., Mannheim, K., 2000, A$\&$A 359, 948

\bibitem[]{i10} Roedig, C., Sesana, A., Dotti, M., Cuadra, J., Amaro-Seoane, P., Haart, F., 
2012, A$\&$A, 545, 127

\bibitem[]{RoeK} Roedig, C. ,  Krolik, J.,  Miller, M.,  2014, ApJ, 785, 115

\bibitem[2014]{Roe} Roedig, C., Sesana, A., 2014, MNRAS, 439, 3476


\bibitem[]{i15} Rossi, E. M., Lodato, G., Armitage, P. J., Pringle, J. E., King, A. R., 2010, MNRAS, 
401, 2021

\bibitem[]{Sch} Scheuer, P. A. G., Feiler, R., 1996, MNRAS. 282, 291

\bibitem[]{Sesana} Sesana, A.,  Roedig, C.,  Reynolds, M.,  Dotti, M.,  2012,  MNRAS, 420, 860

\bibitem[]{S} Shakura, N. I., 1973, Soviet Astronomy, 16, 756

\bibitem[]{SS} Shakura, N. I., Sunyaev, R. A., 1973, A$\&$A, 24, 337

\bibitem[]{i14} Shapiro, S. L., 2010, Phys. Rev. D., 81, 4019

\bibitem[]{sor} Sorathia, K. A., Krolik, J. H., Hawley, J. F. 2013, ApJ, 768, 133

\bibitem[2002]{Tan} Tanaka, H., Takeuchi, T., Ward, W. R., 2002, ApJ, 565, 1257 

\bibitem[]{i11} Tanaka, T., Menou, K., Haiman, Z., 2012, MNRAS, 420, 705

\bibitem[]{Thakar} Thakar, A.  R., Ryden, B. S., 1998, ApJ, 506, 93 

\bibitem[]{i27} Valtonen, M. J., Ciprini, S., Lehto, H. J., 2012, MNRAS, 427, 77

\bibitem[]{yu} Yu, Q., Lu, Y., 2001, A$\&$A, 377, 17 

\bibitem[]{i32} Ziglin, S. L., 1975, Soviet Astronomy Letters, 1, 194


     

\end{thebibliography}
\end{document}